\documentclass[aps,11pt]{revtex4}
\usepackage[colorlinks=true,linkcolor=blue,urlcolor=blue,filecolor=black,citecolor=red,pdfstartview=FitV,pdftitle={},pdfsubject={},
pdfkeywords={},pdfpagemode=None,bookmarksopen=true]{hyperref}
\usepackage{graphicx}%Include figure filespp
\usepackage{epstopdf}%
\usepackage{amsmath}
\usepackage{amsfonts}
\usepackage{amssymb}
\usepackage{color}%
\usepackage{dcolumn}% Align table columns on decimal pointppp
\usepackage{slashed}
\usepackage{amssymb,ulem}
\usepackage{float}
\usepackage{amsthm,amsmath,amssymb}
\usepackage{mathrsfs}
\usepackage{subfigure}
\usepackage{wrapfig}
\usepackage{indentfirst}
\usepackage{multirow}
\usepackage{appendix}
\makeatletter

\newcommand{\Rmnum}[1]{\expandafter\@slowromancap\romannumeral #1@}

\makeatother

\begin{document}
\title{Charged dilatonic black holes in dilaton-massive gravity}
\author{Lina Zhang$^{1,2}$\footnote{linazhang@hunnu.edu.cn;},  Qiyuan Pan$^{1,2}$\footnote{Corresponding author: panqiyuan@hunnu.edu.cn;}, Bo Liu$^{3,4}$\footnote{fenxiao2001@163.com;},
Ming Zhang$^{5}$\footnote{mingzhang0807@126.com}, De-Cheng Zou$^{6}$\footnote{Corresponding author: dczou@jxnu.edu.cn}}

\affiliation{$^{1}$Key Laboratory of Low Dimensional Quantum Structures and Quantum Control of Ministry of Education, Synergetic Innovation Center for Quantum Effects and Applications, and Department of Physics, Hunan Normal University, Changsha, Hunan 410081, China\\
$^{2}$Institute of Interdisciplinary Studies, Hunan Normal University, Changsha, Hunan 410081, China\\
$^{3}$School of Physics, Northwest University, Xi'an, 710127, China\\
$^{4}$School of Arts and Sciences, Shaanxi University of Science and Technology, Xi'an, 710021, China\\
$^{5}$Faculty of Science, Xi'an Aeronautical University, Xi'an 710077, China\\
$^{6}$College of Physics and Communication Electronics, Jiangxi Normal  University, Nanchang 330022, China 
}

\date{\today}

\begin{abstract}
\indent

In this paper, we focus on massive Einstein-dilaton gravity including the coupling of dilaton scalar field to massive graviton terms, and then derive static and spherically symmetric solutions of charged dilatonic black holes in four dimensional spacetime. We find that the dilatonic black hole could possess different horizon structures for some suitably  parameters.
%Then, we also investigate the thermodynamic properties of charged dilatonic black holes with single horizon or double horizons, respectively.
Then, we also investigate the thermodynamic properties of charged dilatonic black holes where $f(r)$  approaches $+\infty$ and $-\infty$, respectively.
\end{abstract}

\maketitle

%%%%%%%%%%%%%%%%%%%%%%%%%%%%%%%%%%%%%%%%%%%%%%%%%%%%%%%%%%%%%%%%%%%%%%%%%%%
\section{Introduction}
\label{1s}

Despite Einstein's general relativity (GR) successfully explaining many observations, alternative theories are being sought due to  the cosmological constant problem \cite{weinberg:1989}, and the origin of acceleration of our universe based on the supernova data \cite{Riess:1998,Perlmutter:1999} and cosmic microwave background
(CMB) radiation \cite{Planck:2015fie,WMAP:2003elm}. An alternative theory to GR is dilaton gravity, which arises from the low-energy limit of string theory. In this theory, Einstein's gravity is reinstated alongside a scalar dilaton field through nonminimal coupling with other fields such as axion and gauge fields \cite{Green:1987sw}.
The dilaton field is essential in string theory when coupled with gravity and other gauge fields. Many attempts have been made to investigate this theory.
For instance, Refs.~\cite{Mignemi:1991wa,Poletti:1994ff,Poletti:1994ww,Cai:1997ii,Clement:2002mb,Dehghani:2004sa,Sheykhi:2007wg} found that the dilaton field alters the causal structure of black holes, leading to curvature singularities at finite radii. The dilaton potential can be seen as a generalization of the cosmological constant and can also modify the asymptotic behavior of the solutions.
Refs. \cite{Gao:2004tu,Gao:2004tv} investigated black hole solutions in (A)dS spacetimes by combining three Liouville-type dilaton potentials.
Additionally, scalar-tensor generalizations of GR have been investigated, incorporating various curvature corrections to the standard Einstein-Hilbert Lagrangian coupled with the dilaton scalar field \cite{Berti:2015itd,Pani:2011xm,Yunes:2011we}.
A specific model, Einstein-dilaton-Gauss-Bonnet (EdGB) gravity, was extensively studied in Refs.~\cite{Kanti:1995vq,Torii:1996yi}. It was found that the scalar dilaton acts as secondary hair, with its charge expressed in terms of the black hole mass.
Later, EdGB gravity was used to study black holes in various dimensions \cite{Guo:2008hf,Ohta:2009tb,Ohta:2009pe},
rotating black holes \cite{Kleihaus:2011tg,Maselli:2015tta},
wormholes \cite{Kanti:2011jz},
and rapidly rotating neutron stars \cite{Kleihaus:2016dui}.

From the perspective of modern particle physics \cite{Weinberg:1965pg,Boulware:1975cg}, the
gravity field is described by a unique theory involving a spin-2 graviton.
Massive gravity is a straightforward modification achieved by giving mass to the graviton. Dating back to 1939, Fierz and Pauli \cite{Fierz:1939} formulated a linear theory of massive gravity. However, this theory consistently faces the Boulware-Deser (BD) ghost issue at the nonlinear level \cite{Boulware:1972if,Boulware:1972fr}.
Notice that the authors \cite{Rham:2011tl} of eliminated the BD ghost by introducing higher-order interaction terms into the Lagrangian.
Subsequently, the ghost-free massive theory known as de Rham-Gabadadze-Tolley (dRGT) massive gravity was developed and discussed in Refs.~\cite{Hinterbichler:2012, Rham:2014mg}.
In dRGT massive gravity, (charged) black hole solutions and their thermodynamics in asymptotically AdS spacetime were investigated \cite{Vegh:2013,Babichev:2014ac,Ghosh:2015cva,Cai:2015tb,Xu:2015rfa,Alfredo:2021,Zangeneh:2017,Zou:2016sab,Zou:2017juz}. It was found that the coefficients in the potential associated with the graviton mass play a role similar to that of charge in thermodynamic phase space.
Other black hole solutions were also studied in massive gravity  \cite{Babichev:2014rb,Hendi:2015eb,Zhang:2017lhl,Acena:2018is,Kodama:2014ss,Hendi:2016jx,Hendi:2018td,Hendi:2018gb,Liu:2024}.
Meanwhile, some charged dilatonic black hole solutions have been discovered \cite{Rakhmanov:1994,Poletti:1995,Gao:2004,Hendi:2016MF}.
Recently, Quasi-dilaton massive gravity, a scalar extension of dRGT massive gravity with a shift symmetry, has also been studied \cite{DAmico:2012hia,Gannouji:2013rwa,Wu:2016jfw,Martinovic:2019hpo,Akbarieh:2021vhv}.
Building upon these studies, we aim to extend our investigation by considering the nonminimal coupling of the dilaton field to the graviton, and derive analytical solutions for charged dilatonic black holes in massive dilaton gravity.

The work is organized as follows. In Sec.~\ref{2s}, we will present the static and spherically symmetric charged black hole solutions in four-dimensional massive Einstein-dilaton gravity and investigate the solution structures of charged dilatonic black holes. In Sec.~\ref{3s}, we will discuss the thermodynamics of theses charged dilatonic black holes in
dilaton-massive gravity.
Finally, we close the work with discussions and conclusions in Sec.~\ref{4s}
%%%%%%%%%%%%%%%%%%%%%%%%%%%%%%%%%%%%%%%%%%%%%%%%%%%%%%%%%%%%%%%%%%%%%%%%%%%%%%%%%%%%%%%%%%%%%%%%%%%%

\section{Solutions of Charged dilatonic black holes}
\label{2s}

The action for massive gravity with a nonminimal coupling of dilaton field $\varphi$ in four dimensional spacetime is given by
\begin{equation}\label{eq4:action}
  I=\frac{1}{16 \pi} \int d^{4} x \sqrt{-g}
  \Big[\mathcal {R}  -2(\nabla \varphi)^2 -e^{-2\alpha\varphi}F_{\mu\nu}F^{\mu\nu}
  + m_0^2 \sum_{i=1}^{4} c_i e^{\alpha_{i}\varphi} \mathcal {U}_i(g,h)\Big],
\end{equation}
where $\varphi=\varphi(r)$ is the dilaton scalar field.
The last term in the action denotes the general form of nonminimal coupling between the scalar field and the massive graviton with coupling constants $\alpha_i$. Here $m_0$ is the mass of graviton, and $c_i$ are the number of dimensionless coupling coefficients.
Moreover, $\mathcal{U}_i$ are symmetric polynomials of the eigenvalues of the ~$4\times 4$\ matrix
$K^{\mu}_{~\nu}=\sqrt{g^{\mu\alpha}h_{\alpha\nu}}$ in which $h$ is a fixed rank-2 symmetric tensor, satisfying the following recursion relation~\cite{Hinterbichler:2012}
\begin{eqnarray}\label{eq4:u1234}
  \mathcal{U}_1& =& [K]=K^{\mu}_{~\mu}, \nonumber\\
  \mathcal{U}_2 &=& [K]^2 - [K^2], \nonumber\\
  \mathcal{U}_3 &=& [K]^3 - 3[K][K^2] + 2[K^3],\\
  \mathcal{U}_4 &=& [K]^4 - 6[K^2][K]^2 + 8[K^3][K] + 3[K^2]^2 - 6[K^4].\nonumber
  \end{eqnarray}
Varying the action with respect to the field variables $g_{\mu\nu}$ and $\varphi$, the equations of motion are obtained as
\begin{eqnarray}
  G_{\mu\nu}&=& \mathcal {R}_{\mu\nu}-\frac{1}{2} \mathcal {R}g_{\mu\nu}=
  2e^{-2\alpha\varphi}[F_{\mu\eta}F^{\eta}_{\nu}-\frac{1}{4}g_{\mu\nu}F_{\eta\tau}F^{\eta\tau}]
  +2\partial_{\mu}\varphi\partial_{\nu}\varphi  - (\nabla\varphi)^2 g_{\mu\nu}+m_0^2 \chi_{\mu\nu}, \label{eq4:eist} \\
  \nabla^2\varphi  &=& -\frac{1}{4}[2\alpha e^{-2\alpha\varphi} F_{\eta\tau}F^{\eta\tau} +m_0^2 \sum_{i=1}^{4} \frac{\partial \tilde{c}_i}{\partial \varphi} \mathcal {U}_i],\label{eq4:dilaton}\\
  &&\nabla_{\mu}(e^{-2\alpha\varphi}F^{\mu\nu})  = 0,\label{eq4:magnetic}
\end{eqnarray}
where
\begin{eqnarray}
  \tilde{c}_i&=&c_i e^{\alpha_{i}\varphi},\\ \nonumber
  \chi_{\mu\nu}&=& \frac{\tilde{c}_1}{2}(\mathcal{U}_1 g_{\mu\nu}-K_{\mu\nu})
  +\frac{\tilde{c}_2}{2}(\mathcal{U}_2 g_{\mu\nu}-2\mathcal{U}_1 K_{\mu\nu}
  +2K^2_{\mu\nu})\nonumber\\
  &+& \frac{\tilde{c}_3}{2}(\mathcal{U}_3 g_{\mu\nu}-3\mathcal{U}_2 K_{\mu\nu}
  +6\mathcal{U}_1 K^2_{\mu\nu}-6K^3_{\mu\nu})\nonumber \\
  &+& \frac{\tilde{c}_4}{2}(\mathcal{U}_4 g_{\mu\nu}-4\mathcal{U}_3 K_{\mu\nu}
  +12\mathcal{U}_2 K^2_{\mu\nu}-24\mathcal{U}_1 K^3_{\mu\nu}
  +24K^4_{\mu\nu})\label{eq4:chi}.
\end{eqnarray}

Now we introduce the static and spherical symmetry metric ansatz
\begin{eqnarray}\label{eq4:metric}
  ds^2 = -f(r)  dt^2 +\frac{dr^2 }{f(r)}+ r^2 R(r)^2 d\Omega^2,
\end{eqnarray}
in which $f(r)$ and $R(r)$ are functions of $r$ and $d\Omega^2=d\theta^2+\sin^2\theta d\phi^2$\ is the line element for the two dimensional spherical  subspace  with the constant curvature.
The fiducial metric $h_{\mu\nu}$ in the action \eqref{eq4:action} serves as a Lagrange multiplier to eliminate the BD ghost~\cite{Vegh:2013}. Choosing an appropriate form simplifies the calculation. Ref.~\cite{Vegh:2013} noted that, unlike the dynamical physical metric $g_{\mu\nu}$, the reference metric $h_{\mu\nu}$ is typically fixed and assumed to be non-dynamical in the massive theory.
In this work, we will follow \cite{Cai:2015tb,Xu:2015rfa} by choosing the fiducial metric to be
\begin{eqnarray}\label{eq4:ref metr}
h_{\mu\nu}=diag(0, 0, c_0^2, c_0^2\sin^2\theta).
\end{eqnarray}
From the ansatz (\ref{eq4:ref metr}), the interaction potential in Eq.~(\ref{eq4:u1234}) changes into
\begin{eqnarray}\label{u1:u2}
\mathcal{U}_1=\frac{2 c_0}{r R},\ \ \
\mathcal{U}_2=\frac{2c_0^2}{ r^2 R^2 },\ \ \
\mathcal{U}_3=\mathcal{U}_4=0.
\end{eqnarray}
From integrating the Maxwell equation (\ref{eq4:magnetic}), the electromagnetic field tensor can be given
\begin{eqnarray}\label{Ftr}
F_{t r}=\frac{Q e^{2\alpha\varphi} }{r^2 R^2 },
\end{eqnarray}
where $Q$ is an integration constant related to the electric charge of the black hole.
Then, $\chi^{\mu}_{~\nu}$ from Eq. (\ref{eq4:chi}) becomes
\begin{eqnarray}\label{eq4:chi:1}
\chi^1_{~1} = \chi^2_{~2} = \frac{c_0(c_1 r R e^{\alpha_1 \varphi }   + c_0 c_2 e^{\alpha_2 \varphi} ) }{r^2 R^2 },\ \ \
\chi^3_{~3} = \chi^4_{~4} = \frac{c_0 c_1 e^{\alpha_1 \varphi }}{2 r R  },
\end{eqnarray}
and the corresponding components of the equation of motion (\ref{eq4:eist}) can be simplified to
\begin{flalign}
G^1_{~1}&=\frac{ -1 + r R  f' (R + r R') + f \left[ R^2 + r^2 R'^2 + 2 r R \left( 3 R' + r R'' \right) \right]}{{r^2 R^2}}\nonumber\\
&=-f \varphi'^2+m_0^2 \chi^1_{~1}-\frac{Q^2 e^{2 \alpha \varphi} }{r^4 R^4}
\label{eq4:inseq:11},\\
G^2_{~2}&=\frac{{-1 +r R f' (R +r R' ) + f (R +r R')^2}}{{r^2 R^2}}
= f \varphi'^2+m_0^2 \chi^2_{~2}-\frac{Q^2 e^{2 \alpha \varphi}}{r^4 R^4}
\label{eq4:inseq:22},\\
G^3_{~3}&=G^4_{~4}=\frac{1}{{2 r R}} \left[2 \left( 2 f + r f' \right) R' + R \left( 2 f' + r f'' \right) + 2 r f R'' \right]
 \nonumber \\
&=-f \varphi'^2+m_0^2 \chi^3_{~3}+\frac{Q^2 e^{2 \alpha \varphi} }{r^4 R^4}
\label{eq4:inseq:33}.
\end{flalign}
Here the prime $'$ denotes differentiation with respect to the radial coordinate $r$.

Based on Eqs.\eqref{eq4:chi:1}-\eqref{eq4:inseq:22},
 we obtain
 \begin{equation}\label{eq4:inseq:12}
 2R'(r) + r \left[R(r) \varphi'(r)^2 + R''(r)\right]
=0.
 \end{equation}
We assume that the dilaton field can be expressed as
\begin{equation}\label{eq4:R(r)}
 R(r) = e^{\beta \varphi(r)},
\end{equation}
where $\beta $ is a constant. 
In fact, the similar assumption (\ref{eq4:R(r)}) has been extensively used to look for the charged dilaton black hole solutions \cite{Alfredo:2021,Zangeneh:2017} in the Maxwell-dilaton gravity. By solving Eq. \eqref{eq4:inseq:12}, we obtain the dilaton field as
\begin{equation}\label{eq5:dilaton field}
  \varphi (r) =-\frac{\beta  }{1 + \beta^2}\ln(\frac{a r}{1+b r}),
\end{equation}
where $a$ and $b$ are integration constants, and we set $a=1$ and $b=0$.

 Considering the scalar field equation \eqref{eq4:dilaton} and substitute the metric ansatz \eqref{eq4:metric},  the scalar field equation becomes
\begin{eqnarray}
 &&c_0^2 e^{\alpha_2 \varphi(r)} m_0^2 \alpha_2 c_2 - \frac{2 e^{2 \alpha \varphi(r)} Q^2 \alpha}{r^2 R(r)^2} + r R(r) c_0 e^{\alpha_1 \varphi(r)} m_0^2 \alpha_1 c_1 +2\frac{d}{dr} \left[ r^2 R(r)^2 f(r) \varphi'(r) \right] = 0.
 \label{eq4:dilaton:2}
\end{eqnarray}
According to Eq. (\ref{eq4:dilaton:2}), along with the assumption $R(r)$ (\ref{eq4:R(r)}) and the dilaton field $\varphi (r)$ (\ref{eq5:dilaton field}), we can solve for $f(r)$ as
\begin{eqnarray}\label{eq4:metr fina}
%%%%%%%%%%%%%%
 f(r) &=& -2 m r^{1 - \frac{2}{1 + \beta^2}} + \frac{Q^2  \alpha (1 + \beta^2)^2 r^{2 - \frac{2(2 + \alpha \beta)}{1 + \beta^2}}}{\beta (1 + 2 \alpha \beta - \beta^2)} \nonumber\\
 &+& \frac{c_0 m_0^2 (1 + \beta^2)^2 c_1 \alpha_1  r^{\frac{1 + 2 \beta^2 - \beta \alpha_1}{1 + \beta^2}}}{2 \beta (2 + \beta^2 - \beta \alpha_1)} + \frac{c_0^2 m_0^2  (1 + \beta^2)^2 c_2 \alpha_2 r^{\frac{\beta (2 \beta - \alpha_2)}{1 + \beta^2}}}{2 \beta (1 + \beta^2 - \beta \alpha_2)},
\end{eqnarray}
where $m$ is an integration constant related to the mass of the black hole.

On the other hand, we further consider the $G^1_{~1}$ component of the gravitational field equation, Eq. (\ref{eq4:inseq:11}) can be expressed as
\begin{eqnarray}
 &&e^{2 \alpha \varphi(r)} Q^2 + r^2 R(r)^2 \left[r^2 f(r) R'(r)^2-1 - c_0^2 e^{\alpha_2 \varphi(r)} m_0^2 c_2 \right] + r^2 R(r)^4 \left[f(r) + r f'(r) + r^2 f(r) \varphi'(r)^2\right] \nonumber\\
 &&+ r^3 R(r)^3 \{r f'(r) R'(r) + 2 f(r) \left[3 R'(r) + r R''(r)\right]-c_0 e^{\alpha_1 \varphi(r)} m_0^2 c_1 \}
 = 0.
  \label{eq4:inseq:33:2}
\end{eqnarray}
Substituting Eq. (\ref{eq4:metr fina}) into Eq. (\ref{eq4:inseq:33:2}), the following parameter constraints can be obtained
\begin{eqnarray}\label{eq:parameters}
%%%%%%%%%%%%%%
 \beta = \alpha,  \quad \alpha_1 = \frac{1}{\alpha}, \quad \alpha_2 = 2 \alpha, \quad \alpha = m_0\sqrt{\frac{c_0 c_1}{2 + 2 c_0 m_0^2 c_1}}.
\end{eqnarray}
It should be noted that for a real $\alpha$, it is required that $m_0^2 c_0 c_1 > 0$ or $m_0^2 c_0 c_1 < -1$, thereby excluding the range $-1 < m_0^2 c_0 c_1 < 0$.

Through (\ref{eq4:metr fina}) and (\ref{eq:parameters}), we can obtain the final solution as
\begin{eqnarray}\label{eq5:metr zuizhong}
%%%%%%%%%%%%%%
f(r) &=& -2m r^{-\frac{1}{3} - \frac{4}{6 + 9 c_0 m_0^2 c_1}}  + \frac{Q^2  (2 + 3 c_0 m_0^2 c_1) r^{-\frac{4}{3} - \frac{4}{6 + 9 c_0 m_0^2 c_1}}}{2 (1 + c_0 m_0^2 c_1)}\nonumber\\
&+& \frac{1}{2}  (2 + 3 c_0 m_0^2 c_1)r^{\frac{2 c_0 m_0^2 c_1}{2 + 3 c_0 m_0^2 c_1}} + \frac{c_0^2 m_0^2 (2 + 3 c_0 m_0^2 c_1)^2 c_2}{2 (1 + c_0 m_0^2 c_1) (2 + c_0 m_0^2 c_1)}.
\end{eqnarray}
Our results reduce to the Schwarzschild case when $c_1 = c_2 = Q = 0$, and reduce to the Reissner-Nordstr\"{o}m case when $c_1 = c_2 = 0$.

In order to comprehend the behaviour of the metric function deeply, we would like to give graphical dependence of the function $f(r)$, and we set  $m_0=c_0=1$ for simplify in following discussions. Then, Eq.~(\ref{eq5:metr zuizhong}) becomes
\begin{eqnarray}\label{eq:simple}
f(r) = -2m r^{-\frac{2 + c_1}{2 + 3 c_1}}  + \frac{Q^2 (2 + 3 c_1)r^{-\frac{4 (1 + c_1)}{2 + 3 c_1}} }{2 (1 + c_1)} + \frac{1}{2}  (2 + 3 c_1) r^{\frac{2 c_1}{2 + 3 c_1}} + \frac{(2 + 3 c_1)^2 c_2}{2 (1 + c_1) (2 + c_1)}.
\end{eqnarray}
Evidently, the parameter $c_1$ affects the asymptotic behavior of $f(r)$. Whereas, $-1 < c_1 < 0$ is excluded.

Now, we calculate the Ricci and Kretschmann scalars to check the spacetime singularities 
\begin{eqnarray}
\mathcal {R}\ \  &=& \frac{2 - r R \left[ 4 f' (R + r R') + r R f'' \right] - 2 f \left[ R^2 + r^2 R'^2 + 2 r R \left( 3 R' + r R'' \right) \right]}{{r^2 R^2}} 
,\\
%\nonumber
  \mathcal {R}^{\mu\nu\rho\sigma} \mathcal {R}_{\mu\nu\rho\sigma}
   &=&\frac{2 r^2 R^2  f'^2 (R + r R')^2 + 4 (f (R + r R')^2 -1 )^2 + r^4 R^4 f''^2 }{r^4 R^4}\nonumber\\
&+&\frac{ 2 r^2 R^2 \left[ 4 f R' + f' (R + r R') + 2 r f R''^2 \right]^2}{r^4 R^4},
\end{eqnarray}
where function $R$ has been shown in Eq.~\eqref{eq4:R(r)}.
One can find that both the Ricci and Kretschmann scalars remain nonsingular at the horizons, indicating that these points are merely coordinate singularities, typical for black holes. Considering the leading terms of asymptotical behaviors of metric at the origin, we obtain
\begin{eqnarray}
  \lim_{r \rightarrow 0} \mathcal {R} &\sim&   r^{-\frac{2}{3} \left(5 + \frac{2}{2 + 3 c_1} \right)}
,  \ \ \ \ \ \ \ \lim_{r\rightarrow0} \mathcal {R}^{\mu\nu\rho\sigma}\mathcal {R}_{\mu\nu\rho\sigma}\sim   r^{-\frac{4}{3} \left(5 + \frac{2}{2 + 3 c_1} \right)}.
\end{eqnarray}
For $c_1 < -1$ or $c_1 > 0$, the Ricci and Kretschmann scalars diverge at the origin $r=0$ but finite for $r>0$. This suggests that the origin $r=0$ is an essential and physical singularity in the spacetime.

\begin{figure}[H]
  \subfigure[]{\label{fig:1:1} %% label for second subfigure
  \includegraphics[width=7cm]{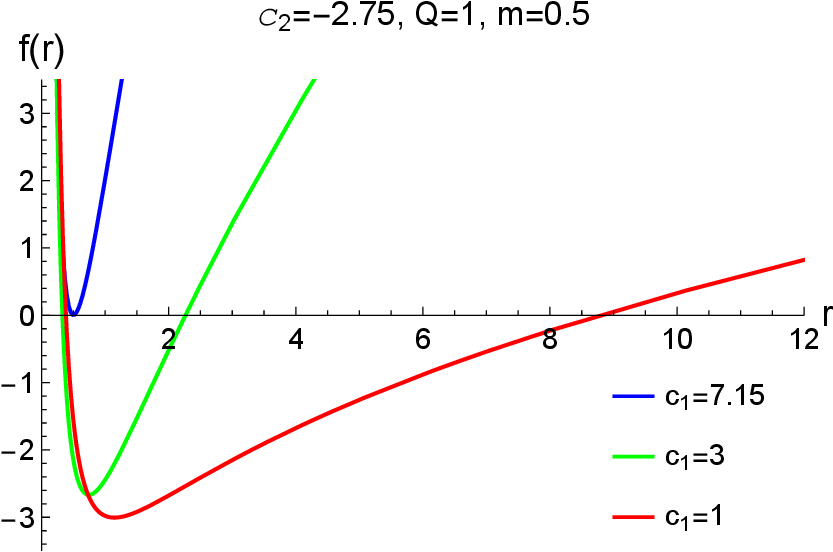}}%\\
   \quad
  \subfigure[]{\label{fig:2:1} %% label for second subfigure
  \includegraphics[width=7cm]{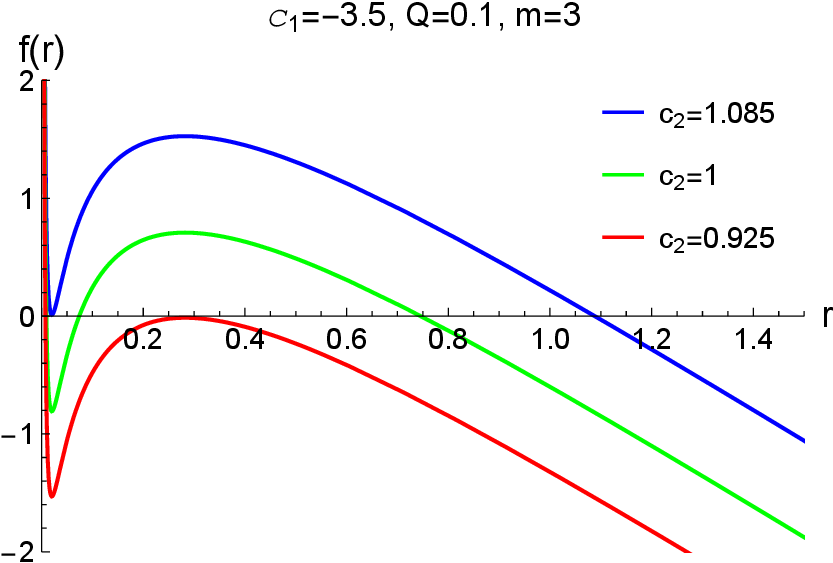}}%\\
  \caption{The function $f(r)$ versus $r$ for different values of $c_1$ and $c_2$.}
    \label{fig:1}
\end{figure}

In order to study the asymptotic behavior of the
solutions, we expand the metric function $f(r)$ for $r \rightarrow \infty$
limit. The figures for metric function $f(r)$ versus radius  $r$ are plotted in Fig. \ref{fig:1}. 
If taking $c_1 < -1 $ or $c_1 > 0$, we have
\begin{eqnarray}\label{eq:wuqio1}
\lim\limits_{r \to \infty}f(r) =  \frac{1}{2}  (2 + 3 c_1) r^{\frac{2 c_1}{2 + 3 c_1}}.
\end{eqnarray}

In case of $c_1 > 0$, the function $f(r)$ approaches $+\infty$ as $r \rightarrow \infty$. As shown in Fig. \ref{fig:1:1}, there are the inner horizon and event horizon when $c_2 = -2.75$, $Q = 1$, $m = 0.5$, and $c_1 = 1$. A naked singularity may appear with the increasing of $c_1$.

In case of $c_1 < -1$, $f(r)$ approaches $-\infty$ as $r \rightarrow \infty$.
Fig. \ref{fig:2:1} shows a complicated black hole spacetime with $c_1 = -3.5$, $Q = 0.1$, $m = 3$, and $c_2 = 1$, where three horizons emerge: the inner and outer event horizons, and the cosmological horizon.
As $c_2$ increases further to a certain value, the inner and outer event horizons coincide.
As $c_2$ decreases further to a certain value, an extremal black hole known as the Nariai black hole may form, exhibiting a coincidence of the event and cosmological horizons.
This indicates that, with the nonminimal coupling to the dilaton field, $c_2$ plays a crucial role in the behavior of the solution $f(r)$.

%%%%%%%%%%%%%%%%%%%%%%%%%%%%%%%%%%%%%%%%%%%%%%%%%%%%%%%%%%%%%%
\section{Thermodynamics of Charged dilatonic black holes}
\label{3s}

%The presence and number of horizons are significantly affected by the parameters of massive gravity, as concluded in the previous Section. The geometrical structure of the black holes depends on the number and type of horizons. By suitable choices of different parameters, they may have: three horizons, two horizons, one horizon and no horizon.

The parameter $c_1$ determines the asymptotic behavior of $f(r)$, as concluded in the previous Section.
For $c_1 > 0$, $f(r)$ approaches $+\infty$.
And for $c_1 < -1$, $f(r)$ approaches $-\infty$.
Now we plan to investigate the thermodynamics of theses charged dilatonic black holes in dilaton-massive gravity.

\subsection{Black holes with $f(r)\rightarrow +\infty$}

The black hole mass can be expressed in terms of the mass parameter $m$ as mentioned before. Considering that the asymptotic behavior of the metric function (\ref{eq:simple}) is unusual, we will use the Brown and York quasilocal formalism to obtain the quasilocal mass $M$ \cite{Brown:1993,Brown:1994}. For the metric (Eq. (2.7) in Ref. \cite{Chan:1995} or Eq. (3.10) in Ref. \cite{Dehghani:2019})
\begin{eqnarray}\label{eq4:metric-2}
  ds^2 = -X(\rho)^2 dt^2 + \frac{d\rho^2}{Y(\rho)^2} + \rho^2 (d\theta^2 + \sin^2 \theta d\phi^2),
\end{eqnarray}
provided that the matter field does not contain derivatives of the metric, we can get the quasilocal black hole mass $M$ by using the following relation (Eq. (2.8) in Ref. \cite{Chan:1995} or Eq. (3.11) in Ref. \cite{Dehghani:2019}):
\begin{equation}\label{eq4:metric-M}
\mathcal{M}(\rho) = \rho X(\rho) \left[Y_0(\rho) - Y(\rho)\right],
\end{equation}
with a background metric function $Y_0(\rho)$ which establishes the zero of the mass, and taking the limit $\rho \rightarrow \infty$:
\begin{equation}\label{eq4:metric-Mm}
M = \lim_{\rho \rightarrow \infty} \mathcal{M}(\rho).
\end{equation}

Now, we write the metric (\ref{eq4:metric}) in the form of Eq. (\ref{eq4:metric-2}) by considering the transformation
$\rho=r R(r)$. Thus, one can show that
\begin{equation}\label{eq4:dr-drho}
dr^2 = \frac{(1+\beta^2)^2}{ R(r)^2} d\rho^2.
\end{equation}
In our case, we find
\begin{align}\label{eq4:XY-f}
X(\rho)^2 = f(r), \quad Y(\rho)^2 = \frac{R(r)^2}{(1 + \beta^2)^2} f(r),
\end{align}
with $r=r(\rho)$. Substituting these quantities into Eq. (\ref{eq4:metric-M}), we obtain
\begin{equation}\label{eq4:M1}
\mathcal{M} = \frac{r R(r)^2}{1+\beta^2}\{[f(r)f_0(r)]^{\frac{1}{2}}-f(r)\},
\end{equation}
in which
\begin{eqnarray}\label{eq:simple-ur}
f_0(r) = \frac{Q^2 (2 + 3 c_1)r^{-\frac{4 (1 + c_1)}{2 + 3 c_1}} }{2 (1 + c_1)} + \frac{1}{2}  (2 + 3 c_1) r^{\frac{2 c_1}{2 + 3 c_1}} + \frac{(2 + 3 c_1)^2 c_2}{2 (1 + c_1) (2 + c_1)},
\end{eqnarray}
where $f_0(r)$ is the metric function $f(r)$ evaluated for the value $m = 0$ of the integration constant.
Now, Eq. (\ref{eq4:M1}) can be rewritten as
\begin{eqnarray}\label{eq4:M1-1}
\mathcal{M} &=& \frac{2 + 2c_1}{2 + 3c_1} r^{\frac{2 + c_1}{2 + 3 c_1}}\Big\{\{[-2m r^{-\frac{2 + c_1}{2 + 3 c_1}}+f_0(r)]f_0(r)\}^{\frac{1}{2}}+2m r^{-\frac{2 + c_1}{2 + 3 c_1}}-f_0(r)\Big\}\nonumber \\
&=&\frac{2 + 2c_1}{2 + 3c_1} r^{\frac{2 + c_1}{2 + 3 c_1}}\Big\{f_0(r)\Big(1+\frac{1}{2}\Big[\frac{-2m r^{-\frac{2 + c_1}{2 + 3 c_1}}}{f_0(r)}\Big]-\frac{1}{8}\Big[\frac{-2m r^{-\frac{2 + c_1}{2 + 3 c_1}}}{f_0(r)}\Big]^2+ \mathcal{O}\Big[\frac{-2m r^{-\frac{2 + c_1}{2 + 3 c_1}}}{f_0(r)}\Big]^3\Big)\nonumber \\
&+&2m r^{-\frac{2 + c_1}{2 + 3 c_1}}-f_0(r)\Big\}.
\end{eqnarray}
which leads to the black hole mass
\begin{eqnarray}\label{eq4:M1-2}
M = \lim_{r \rightarrow \infty} \mathcal{M} = \lim_{r \rightarrow \infty} \frac{2 + 2c_1}{2 + 3c_1}\Big\{m-\frac{m^2}{2}\Big[\frac{ r^{-\frac{2 + c_1}{2 + 3 c_1}}}{f_0(r)}\Big]-\mathcal{O}\Big[\frac{ r^{-\frac{2 + c_1}{2 + 3 c_1}}}{f_0(r)}\Big]^2 \Big\}.
\end{eqnarray}
Note that $f(r)$ approaches $+\infty$ as $r \rightarrow \infty$, which requires $c_1 > 0$. Obviously, it is shown that $\lim_{r \rightarrow \infty} [\frac{ r^{-\frac{2 + c_1}{2 + 3 c_1}}}{f_0(r)} ]$ and its higher powers are equal to zero. As a result, we have
\begin{eqnarray}\label{eq4:mass-1}
  M=\frac{2m (1 + c_1)}{2 + 3 c_1}.
\end{eqnarray}
%Here one can obtain $m<0$ with a positive ADM mass because $c_1> 0$. 
Thus, according to the definition of horizon $f(r_h)=0$, the mass of charged dilatonic black hole is given by
\begin{eqnarray}\label{eq4:mass}
  M&=&\frac{1}{2} \left[ \frac{Q^2}{r_h} +  \left( 1 + c_1 \right)r_h + \frac{(2 + 3 c_1) c_2 r_h^{\frac{1}{3} + \frac{4}{6 + 9  c_1}} }{2 +  c_1} \right].
\end{eqnarray}

\begin{figure}[H]
  \subfigure[]{\label{fig:m1:1} %% label for second subfigure
  \includegraphics[width=5.3cm]{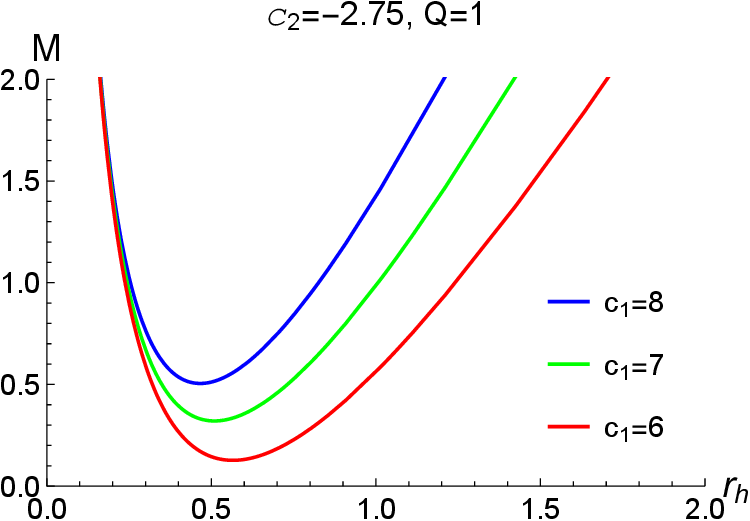}}%\\
  %\hfill%
  \quad
  \subfigure[]{\label{fig:m1:2} %% label for second subfigure
  \includegraphics[width=5.3cm]{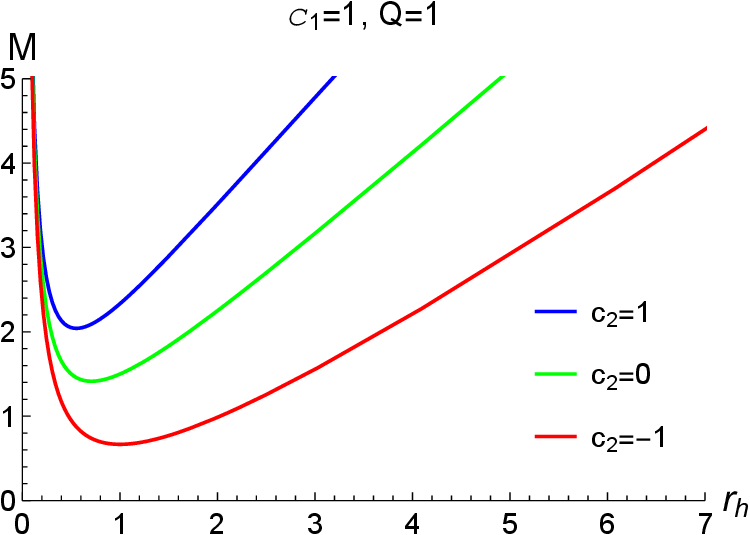}}
  \quad
  \subfigure[]{\label{fig:m1:3} %% label for second subfigure
  \includegraphics[width=5.3cm]{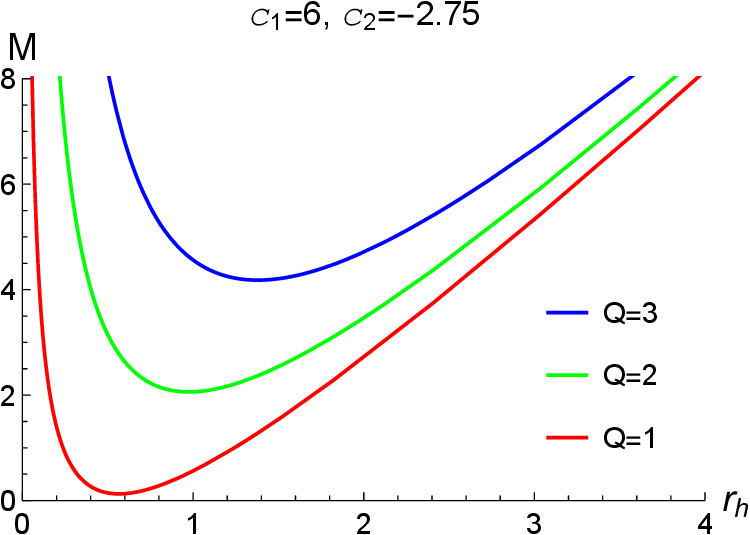}}
  \caption{The mass $M$ of black hole versus $r_h$. }
    \label{fig:m1}
\end{figure}

The quasilocal mass $M$ as a function of black hole radius is plotted in Fig. \ref{fig:m1:1} for various values of $c_1$, Fig. \ref{fig:m1:2} for various values of $c_2$, and Fig. \ref{fig:m1:3} for various values of $Q$.
In all cases, there exists a minimum mass $M_{min}$, and there are two black holes with the same mass, distinguished by their sizes (a smaller one and a larger one). The quasilocal mass $M$ is always positive. It can be observed that the minimum mass $M_{min}$ increases with $c_1$, $c_2$, $Q$, according to the case.

To develop thermodynamics of charged dilatonic black hole, we need to calculate the Hawking temperature of the black hole geometrically associated with the black hole horizon. In terms of the surface gravity $\kappa$ corresponding to  the null killing
vector $(\frac{\partial}{\partial t})^a$  at the horizon, the temperature can be written as
\begin{eqnarray}
T &=& \frac{\kappa}{2 \pi}=\frac{1}{4 \pi}\frac{\partial f(r)}{\partial r}\big|_{r=r_h}= \frac{ (2 + 3  c_1) \left[ -Q^2 + r_h \left(r_h +   c_1 r_h+   c_2 r_h^{\frac{1}{3} + \frac{4}{6 + 9  c_1}}\right) \right]r_h^{-\frac{7}{3} - \frac{4}{6 + 9  c_1}}}{8\pi \left(1 +   c_1\right)}.
   \label{eq4:temp}
\end{eqnarray}

\begin{figure}[H]
  \subfigure[]{\label{fig:t1:1} %% label for second subfigure
  \includegraphics[width=5.3cm]{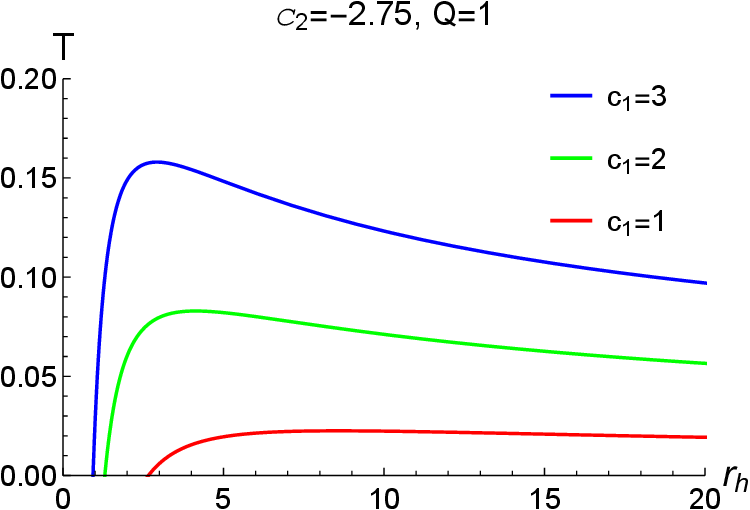}}%\\
    \quad
  \subfigure[]{\label{fig:t1:2} %% label for second subfigure
  \includegraphics[width=5.3cm]{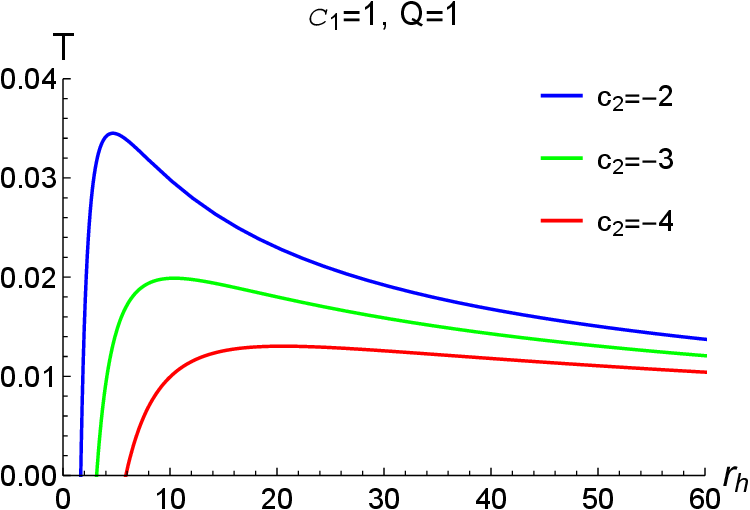}}
   \quad
  \subfigure[]{\label{fig:t1:3} %% label for second subfigure
  \includegraphics[width=5.3cm]{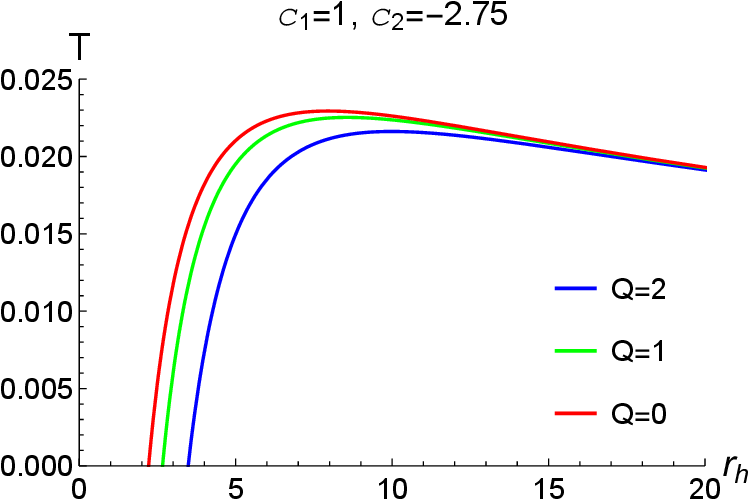}}
  \caption{The temperature $T$ versus $r_h$. }
  (a)The extreme points of each $T-r_h$ curves ($r_{cri}$, $T_{max}$) are (2.92518, 0.157982), (4.12896, 0.0829192), (8.56845, 0.0225279) for $c_1 = 3, 2, 1$;
  (b)The extreme points of each $T-r_h$ curves ($r_{cri}$, $T_{max}$) are (4.66247, 0.034502), (10.3954, 0.0198907), (20.5486, 0.0130369) for $c_2 = -2, -3, -4$;
  (c) The extreme points of each $T-r_h$ curves ($r_{cri}$, $T_{max}$) are (9.98137, 0.0216182), (8.56845, 0.0225279), (7.95021, 0.0229384) for $Q = 2, 1, 0$.
  \label{fig:t1}
\end{figure}

The temperature of the black hole as a function of radius is depicted in Fig. \ref{fig:t1:1} for various values of $c_1$, in Fig. \ref{fig:t1:2} for various values of $c_2$, and in Fig. \ref{fig:t1:3} for various values of $Q$.
In all scenarios, a maximum positive temperature $T_{max}$ is observed for the black hole, indicating the existence of two black holes with identical temperatures but different sizes (a smaller one and a larger one). Moreover, $T_{max}$ increases with $ c_1 $, $ c_2 $, and $ Q $, depending on the case. And there exists a minimum radius $r_{min}$ where the temperature is 0.

The entropy of charged dilatonic black hole is given by
\begin{eqnarray}  \label{eq4:entropy}
S=\pi r_h^2R(r_h)^2=\pi r_h^{\frac{4}{3} \left(1 + \frac{1}{2 + 3  c_1}\right)}.
\end{eqnarray}
Refs.~\cite{Babichev:2014ac}-\cite{Zou:2016sab} have pointed out that the graviton mass does not significantly affect the form of
the entropy, and contributes only as a correction for the horizon radius. 
Then,the potential of charge is
\begin{eqnarray}\label{eq4: U}
  U =\frac{Q}{r_h}.
\end{eqnarray}
It is a matter of calculation to show that the intensive
parameters $T$ and $U$, conjugate to the black hole entropy
and charge, satisfy the following relations:
\begin{eqnarray}\label{eq4: first law-1}
\Bigg(\frac{\partial M}{\partial S} \Bigg)_Q= T,\quad\quad \Bigg(\frac{\partial M}{\partial Q} \Bigg)_S= U.
\end{eqnarray}
Then, we find that the thermodynamic quantities satisfy the first law of black hole thermodynamics
\begin{eqnarray}\label{eq4: first law}
  d M = T d S+U d Q.
\end{eqnarray}

To assess the thermal stability, we compute the heat capacity
\begin{eqnarray}
  C_Q=T \Bigg(\frac{\partial S}{\partial T} \Bigg)_Q= T \Bigg(\frac{\partial S/\partial r_h}{\partial T/\partial r_h}\Bigg)_Q,
  \label{eq4:capcty}
\end{eqnarray}
which leads to
\begin{eqnarray}
C_Q &=& \frac{4 \pi (1 +  c_1) \left[Q^2 - r_h \left(r_h + c_1 r_h  + c_2  r_h^{\frac{1}{3} + \frac{4}{6 + 9  c_1}} \right) \right] r_h^{\frac{4}{3} \left(1 + \frac{1}{2 + 3  c_1}\right)}}{2 r_h^2 - Q^2 \left(6 + 7  c_1\right) +  r_h \left[ c_1\left(3 +  c_1\right)r_h + \left(2 + 3  c_1\right)c_2 r_h^{\frac{1}{3} + \frac{4}{6 + 9  c_1}} \right]}.
  \label{eq4:cap}
\end{eqnarray}

\begin{figure}[H]
  \subfigure[]{\label{fig:cc1:1} %% label for second subfigure
  \includegraphics[width=5.3cm]{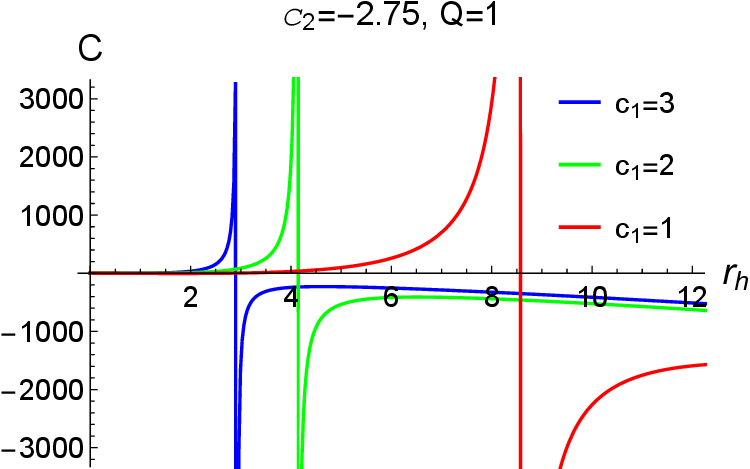}}%\\
 %\hfill
 \quad
  \subfigure[]{\label{fig:cc1:2} %% label for second subfigure
  \includegraphics[width=5.3cm]{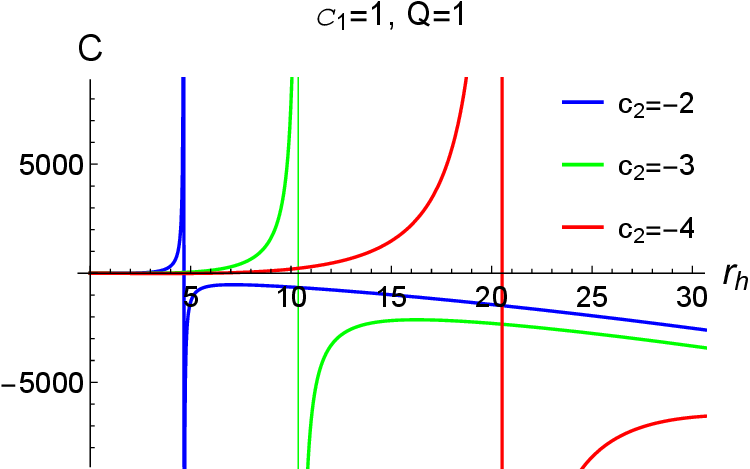}}
  \quad
  \subfigure[]{\label{fig:cc1:3} %% label for second subfigure
  \includegraphics[width=5.3cm]{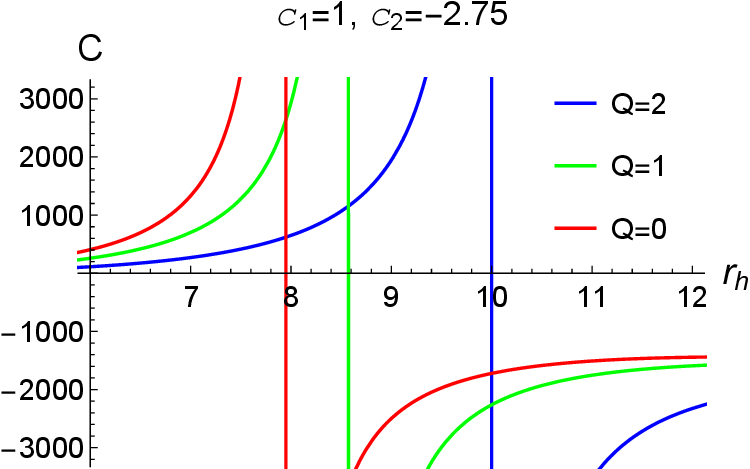}}
  \caption{The  heat capacity $C$ versus $r_h$. }
  \label{fig:cc1}
\end{figure}

One can find that the non-monotonic nature of temperature allows us to
conclude that the heat capacity exhibits discontinuities  as illustrated in Fig.~\ref{fig:cc1}, which shows that black holes with relatively smaller horizon radii ($ r_{min}< r_h< r_{cri}$) are stable thermodynamic systems, while a domain of instability exists for larger radii above the point of discontinuity ($r_h > r_{cri}$).
In other words, the figures clearly illustrate the stable and unstable regions based on the discontinuities observed in the heat capacity as a function of the horizon radius.

\subsection{Black holes with $f(r)$ $\rightarrow$ 
 $-\infty$}

When $f(r)$ is for pure dS space, horizons happen at $f(r) = 0$. The biggest root is the cosmological horizon at $r = r_c$, and the next is the black hole event horizon at $r = r_+$.
The equations $f(r_+) = 0$ and $f(r_c) = 0$ are rearranged to
br as follows \cite{Cai:2002,Ma:2015,Zhang:2019}:
\begin{eqnarray}\label{eq4-1:mass}
  M&=&\frac{2m (1 + c_1)}{2 + 3 c_1}
=\frac{1}{2} \left[ \frac{Q^2}{r_+} +  \left( 1 + c_1 \right)r_+ + \frac{(2 + 3 c_1) c_2 r_+^{\frac{1}{3} + \frac{4}{6 + 9  c_1}} }{2 +  c_1} \right],
\end{eqnarray}
\begin{eqnarray}\label{eq4-2:mass}
  M&=&\frac{2m (1 + c_1)}{2 + 3 c_1}
=\frac{1}{2} \left[ \frac{Q^2}{r_c} +  \left( 1 + c_1 \right)r_c + \frac{(2 + 3 c_1) c_2 r_c^{\frac{1}{3} + \frac{4}{6 + 9  c_1}} }{2 +  c_1} \right],
\end{eqnarray}
from which one can derive
\begin{eqnarray}\label{eq4mr:mass}
  M&=&\frac{ [Q^2 + r_c^2 (1 + c_1)]r_+^{\frac{4 (1 + c_1)}{2 + 3 c_1}} - [Q^2 + r_+^2 (1 + c_1)]r_c^{\frac{4 (1 + c_1)}{2 + 3 c_1}}}{ 2 r_c r_+^{\frac{4 (1 + c_1)}{2 + 3 c_1}}-2 r_+ r_c^{\frac{4 (1 + c_1)}{2 + 3 c_1}} },
\end{eqnarray}
by eliminating the $c_2$, and
\begin{eqnarray}\label{eq4c2:mass}
  c_2&=&\frac{(r_c - r_+) (2 + c_1) [Q^2 - r_c r_+ (1 + c_1)]}{(2 + 3 c_1) r_c r_+ \left(r_c^{\frac{2 + c_1}{2 + 3 c_1}} - r_+^{\frac{2 + c_1}{2 + 3 c_1}}\right) },
\end{eqnarray}
by eliminating the $M$.

The surface gravities of black hole horizon and the cosmological horizon are
\begin{eqnarray}
 T_+&=&\frac{\kappa_+}{2\pi}= \frac{1}{4\pi}\frac{\partial f(r)}{\partial r}\big|_{r=r_+} \nonumber\\
 &=& \frac{ (2 + 3  c_1) \left[ -Q^2 + r_+ \left(r_+ +   c_1 r_+ +   c_2 r_+^{\frac{1}{3} + \frac{4}{6 + 9  c_1}}\right) \right]r_+^{-\frac{7}{3} - \frac{4}{6 + 9  c_1}}}{8\pi \left(1 +   c_1\right)},
   \label{eq4:kh}
\end{eqnarray}
\begin{eqnarray}
 T_c&=&\frac{\kappa_c}{2\pi}= -\frac{1}{4\pi}\frac{\partial f(r)}{\partial r}\big|_{r=r_c} \nonumber\\
 &=& -\frac{ (2 + 3  c_1) \left[ -Q^2 + r_c \left(r_c +   c_1 r_c +   c_2 r_c^{\frac{1}{3} + \frac{4}{6 + 9  c_1}}\right) \right]r_c^{-\frac{7}{3} - \frac{4}{6 + 9  c_1}}}{8\pi \left(1 +   c_1\right)}.
   \label{eq4:kc}
\end{eqnarray}

Considering the connection between the black hole horizon and the cosmological horizon, we can derive the effective thermodynamic quantities and corresponding first law of black hole thermodynamics \cite{Li:2017,Du:2022}:
\begin{eqnarray}  \label{eq4:dm}
dM = T_{eff} dS  + U_{eff} dQ -P_{eff} dV,
\end{eqnarray}
where the effective temperature $T_{eff}$, effective electric potential $U_{eff}$ and the effective pressure $P_{eff}$ are denoted as
\begin{eqnarray}\label{Teff-1}
T_{eff}&=&\Bigg(\frac{\partial M}{\partial S} \Bigg)_{Q,V}= \frac{(\frac{\partial M}{\partial x})_{ r_{c}}(\frac{\partial V}{\partial r_c})_{x}-(\frac{\partial V}{\partial x})_{ r_{c}}(\frac{\partial M}{\partial r_c})_{x}}{(\frac{\partial S}{\partial x})_{ r_{c}}(\frac{\partial V}{\partial r_c})_{x}-(\frac{\partial V}{\partial x})_{ r_{c}}(\frac{\partial S}{\partial r_c})_{x}},
\end{eqnarray}
\begin{eqnarray}\label{Ueff-1}
U_{eff}=\Bigg(\frac{\partial M}{\partial Q} \Bigg)_{S,V},
\end{eqnarray}
\begin{eqnarray}\label{Peff-1}
P_{eff}&=&-\Bigg(\frac{\partial M}{\partial V} \Bigg)_{Q,S}= -\frac{(\frac{\partial M}{\partial x})_{ r_{c}}(\frac{\partial S}{\partial r_c})_{x}-(\frac{\partial S}{\partial x})_{ r_{c}}(\frac{\partial M}{\partial r_c})_{x}}{(\frac{\partial V}{\partial x})_{ r_{c}}(\frac{\partial S}{\partial r_c})_{x}-(\frac{\partial S}{\partial x})_{ r_{c}}(\frac{\partial V}{\partial r_c})_{x}}.
\end{eqnarray}
From Eq. (\ref{eq4mr:mass}), one know that
\begin{eqnarray}\label{zm}
  M&=&\frac{[Q^2 + r_c^2 (1 + c_1)]x^{\frac{4 (1 + c_1)}{2 + 3 c_1}} - [Q^2 + r_c^2 x^2 (1 + c_1)]}{ 2 [x^{\frac{4 (1 + c_1)}{2 + 3 c_1}}-  x] r_c },
\end{eqnarray}
where $x = r_+/r_c $.
Here the thermodynamic volume is that between the black hole
horizon and the cosmological horizon, namely
\begin{eqnarray}  \label{Vc}
V&=&V_c - V_+=\frac{4}{3} \pi r_c^3R(r_c)^3-\frac{4}{3} \pi r_+^3R(r_+)^3=\frac{4}{3}\pi (r_c^{2 + \frac{2}{2 + 3c_1}}-r_+^{2 + \frac{2}{2 + 3c_1}})\nonumber\\
&=&\frac{4}{3}\pi r_c^{2 + \frac{2}{2 + 3c_1}}(1-x^{2 + \frac{2}{2 + 3c_1}}).
\end{eqnarray}
The total entropy can be written as
\begin{eqnarray}\label{eS}
  S &=&S_{+} +  S_{c}+  S_{ex}=\pi r_+^2R(r_+)^2 + \pi r_c^2R(r_c)^2+  S_{ex}\nonumber\\
&=&\pi r_c^{\frac{4}{3} \left(1 + \frac{1}{2 + 3  c_1}\right)}[1+ x^{\frac{4}{3} \left(1 + \frac{1}{2 + 3  c_1}\right)}+F(x)] .
\end{eqnarray}
Here the undefined function $F(x)$ represents the extra contribution from the correlations of the two horizons. Then, how to determine the function $F(x)$?

In general, the temperatures of the black hole horizon and the cosmological horizon differ, so the globally effective temperature $T_{eff}$ cannot be compared to them. However, in special cases like the lukewarm case, the temperatures of the two horizons are equal. We conjecture that in this special case, the effective temperature should also be the same. Based on this consideration, we can determine the function $F(x)$.

Substituting Eq. (\ref{zm}), Eq. (\ref{Vc}) and Eq. (\ref{eS}) into Eq. (\ref{Teff-1}), one can obtain
\begin{eqnarray}\label{eq4: first law-22}
T_{eff}&=&\Bigg\{r_c^{-\frac{6 + 7c_1}{2 + 3c_1}} \Big\{Q^2 \Big[-4x^{\frac{{4(1+c_1)}}{{2 + 3c_1}}} (1 + c_1) - 4x^{\frac{{12+13c_1}}{{2 + 3c_1}}} (1 + c_1) + x^{\frac{{10(1+c_1)}}{{2 + 3c_1}}} (2 + c_1) \nonumber\\
&+& x^{\frac{{6 + 7c_1}}{{2 + 3c_1}}} (2 + c_1) + x(2 + 3c_1) + x^{\frac{{14(1+c_1)}}{{2 + 3c_1}}} (2 + 3c_1)\Big] + r_c^2 \Big[2x^{\frac{{8 + 10c_1}}{{2 + 3c_1}}} c_1 (1 + c_1)  \nonumber\\
&+& 2x^{\frac{{12 + 13c_1}}{{2 + 3c_1}}} c_1 (1 + c_1) + x^{\frac{{6 + 7c_1}}{{2 + 3c_1}}} (2 + 3c_1 + c_1^2) + x^{\frac{{2(7 + 8c_1)}}{{2 + 3c_1}}} (2 + 3c_1 + c_1^2) - x^3 (2 + 5c_1 + 3c_1^2) \nonumber\\
&-& x^{\frac{{14(1+c_1)}}{{2 + 3c_1}}} (2 + 5c_1 + 3c_1^2) \Big] \Big\}\Bigg\}/ \Big\{2\pi x [x - x^{\frac{{4(1+c_1)}}{{2 + 3c_1}}}]^2 \Big[ -4x^{\frac{2}{{2 + 3c_1}}} (x + x^{\frac{{c_1}}{{2 + 3c_1}}}) (1 + c_1) \nonumber\\
&-& 4x^{1 + \frac{2}{{2 + 3c_1}}} (1 + c_1) F(x) + ( x^{2 + \frac{2}{{2 + 3c_1}}}-1 ) (2 + 3c_1) F'(x)\Big] \Big\},
\end{eqnarray}
substituting Eq. (\ref{zm}) into Eq. (\ref{Ueff-1}), one can obtain
\begin{eqnarray}\label{eq4: first law-11}
U_{eff}=\frac{Q [ x^{\frac{4 (1 + c_1)}{2 + 3 c_1}}-1 ]}{r_c [ x^{\frac{4 (1 + c_1)}{2 + 3 c_1}}-x ]},
\end{eqnarray}
substituting Eq. (\ref{zm}), Eq. (\ref{Vc}) and Eq. (\ref{eS}) into Eq. (\ref{Peff-1}), one can obtain
\begin{eqnarray}\label{eq4: first law-44}
P_{eff}&=&\Bigg\{r_c^{-\frac{8 + 9c_1}{2 + 3c_1}} \Big\{ -12(1 + c_1) \Bigg[\Big[ 2 + 3c_1 - 4x^{\frac{2 + c_1}{2 + 3c_1}}(1 + c_1) \Big] \Big[ Q^2 + r_c^2x^2(1 + c_1) - x^{\frac{4(1 + c_1)}{2 + 3c_1}}[Q^2 \nonumber\\
&+& r_c^2(1 + c_1)] \Big] - 2[x - x^{\frac{4(1 + c_1)}{2 + 3c_1}}]\Big[r_c^2x(1 + c_1)(2 + 3c_1)- 2x^{\frac{2 + c_1}{2 + 3c_1}}(1 + c_1)[Q^2 + r_c^2(1 + c_1)]\Big] \Bigg]\nonumber\\
&\times&\Big[1 + x^{\frac{4(1 + c_1)}{2 + 3c_1}} + F(x)\Big] + [x - x^{\frac{4(1 + c_1)}{2 + 3c_1}}](2 + 3c_1)\Big[ r_c^2[x^{\frac{4(1 + c_1)}{2 + 3c_1}}-x^2 ](1 + c_1) -Q^2[ x^{\frac{4(1 + c_1)}{2 + 3c_1}}-1 ]\Big]\nonumber\\
&\times&\Big[12x^{\frac{2 + c_1}{2 + 3c_1}}(1 + c_1) + 3(2 + 3c_1) F'(x)\Big] \Big\}\Bigg\}/\Bigg\{48\pi[x - x^{\frac{4(1 + c_1)}{2 + 3c_1}}]^2(1 + c_1)\Big[4x^{\frac{2}{2 + 3c_1}}\nonumber\\
&\times&(x + x^{\frac{c_1}{2 + 3c_1}})(1 + c_1) + 4x^{1 + \frac{2}{{2 + 3c_1}}}(1 + c_1) F(x)- (x^{2 + \frac{2}{2 + 3c_1}}-1 )(2 + 3c_1) F'(x)\Big]\Bigg\}.
\end{eqnarray}

As is well known, in the lukewarm case, we have $T_+=T_c$ \cite{Zhang:2016,Zhao:2021}, there is 
\begin{eqnarray}\label{Q2}
  Q^2 = \frac{{r_c^2x(1+c_1)\Big[-x^{\frac{{2+c_1}}{{2+3c_1}}}(2+c_1)+x^{\frac{{6+7c_1}}{{2+3c_1}}}(2+c_1)+x(2+3c_1)-x^{\frac{{6+5c_1}}{{2+3c_1}}}(2+3c_1)\Big]}}{{2+3c_1-4x^{\frac{{2+c_1}}{{2+3c_1}}}(1+c_1)+4x^{\frac{{6+7c_1}}{{2+3c_1}}}(1+c_1)-x^{\frac{{8(1+c_1)}}{{2+3c_1}}}(2+3c_1)}}.
\end{eqnarray}
Using this, we can determine the effective temperature of a lukewarm black hole:
\begin{eqnarray}
\label{Teff}
T_{eff} &=& r_c^{-\frac{6+7c_1}{2+3c_1}} \Bigg\{ \Bigg[r_c^2 x(1+c_1) \Big[-4x^{\frac{4(1+c_1)}{2+3c_1}}(1+c_1) - 4x^{\frac{12+13c_1}{2+3c_1}}(1+c_1) + x^{\frac{10(1+c_1)}{2+3c_1}}(2+c_1)\nonumber \\
&+& x^{\frac{6+7c_1}{2+3c_1}}(2+c_1) + x(2+3c_1) + x^{\frac{14(1+c_1)}{2+3c_1}}(2+3c_1)\Big]\Big[-x^{\frac{2+c_1}{2+3c_1}}(2+c_1) + x^{\frac{6+7c_1}{2+3c_1}}(2+c_1)\nonumber \\
&+& x(2+3c_1) - x^{\frac{6+5c_1}{2+3c_1}}(2+3c_1)\Big]\Bigg] \Bigg/ \Bigg[ 2+3c_1 - 4x^{\frac{2+c_1}{2+3c_1}}(1+c_1) + 4x^{\frac{6+7c_1}{2+3c_1}}(1+c_1) \nonumber \\
&-& x^{\frac{8(1+c_1)}{2+3c_1}}(2+3c_1) \Bigg] + r_c^2 \Bigg[ 2x^{\frac{8+10c_1}{2+3c_1}}c_1(1+c_1) + 2x^{\frac{12+13c_1}{2+3c_1}}c_1(1+c_1) + x^{\frac{6+7c_1}{2+3c_1}}(2+3c_1+c_1^2) \nonumber \\
&+& x^{\frac{2(7+8c_1)}{2+3c_1}}(2+3c_1+c_1^2)- x^3(2+5c_1+3c_1^2) - x^{\frac{14(1+c_1)}{2+3c_1}}(2+5c_1+3c_1^2) \Bigg] \Bigg\} \Bigg / \nonumber\\
&& \Bigg\{ 2\pi x[x-x^{\frac{4(1+c_1)}{2+3c_1}}]^2\Bigg[ -4x^{\frac{2}{2+3c_1}}(x+x^{\frac{c_1}{2+3c_1}})(1+c_1) - 4x^{1+\frac{2}{2+3c_1}}(1+c_1)F(x)\nonumber\\
&+& \left(-1+x^{2+\frac{2}{2+3c_1}}\right)(2+3c_1)F'(x) \Bigg] \Bigg\}.
\end{eqnarray}
We also note that for the lukewarm black hole, the temperature is
\begin{eqnarray}\label{TcTh}
&&T_+=T_c =-\Big\{r_c^{-\frac{{2+c_1}}{{2+3c_1}}}(2+3c_1)\Big[2c_1-2x^{\frac{{8(1+c_1)}}{{2+3c_1}}}c_1-4x^2(1+c_1)+4x^{\frac{{2(2+c_1)}}{{2+3c_1}}}(1+c_1)\nonumber\\
&&+2x(2+c_1)-2x^{\frac{{6+5c_1}}{{2+3c_1}}}(2+c_1)-2x^{\frac{{2+c_1}}{{2+3c_1}}}(2+3c_1)+x^{\frac{{6+7c_1}}{{2+3c_1}}}(4+6c_1)\Big]\Big\}/\Big\{8\pi(x^{\frac{{2+c_1}}{{2+3c_1}}}-1)\nonumber\\
&&\times\Big[-2-3c_1+4x^{\frac{{2+c_1}}{{2+3c_1}}}(1+c_1)-4x^{\frac{{6+7c_1}}{{2+3c_1}}}(1+c_1)+x^{\frac{{8(1+c_1)}}{{2+3c_1}}}(2+3c_1)\Big]\Big\}.
\end{eqnarray}
Equating temperatures from Eq. (\ref{Teff}) and Eq. (\ref{TcTh}). When $c_1=-3.5$, we obtain the analytic solution to this equation, which is
\begin{align*}\label{Fx}
F(x) = &\frac{1}{50} \left(1 - x^{\frac{30}{17}}\right)^{\frac{2}{3}} \left[50 D_1 - \frac{1}{1 - x^{\frac{3}{17}} - x^{\frac{30}{17}} + x^{\frac{33}{17}}} \left(1 - x^{\frac{30}{17}}\right)^{\frac{1}{3}} \Bigg(80 - 18 x^{\frac{2}{17}} - 97 x^{\frac{3}{17}} + 3 x^{\frac{5}{17}} + 3 x^{\frac{6}{17}} \right. \\
&\left. + 3 x^{\frac{8}{17}} + 3 x^{\frac{9}{17}} + 3 x^{\frac{11}{17}} + 3 x^{\frac{12}{17}} + 3 x^{\frac{14}{17}} + 3 x^{\frac{15}{17}} + 3 x + 3 x^{\frac{18}{17}} + 3 x^{\frac{20}{17}} + 3 x^{\frac{21}{17}} - 47 x^{\frac{23}{17}} + 3 x^{\frac{24}{17}} \right. \\
&\left. + 3 x^{\frac{26}{17}} + 3 x^{\frac{27}{17}} + 3 x^{\frac{29}{17}} - 27 x^{\frac{30}{17}} + 21 x^{\frac{32}{17}}\Bigg)- 18 x^{\frac{2}{17}} \, _2F_1\left(\frac{1}{15}, \frac{2}{3}, \frac{16}{15}, x^{\frac{30}{17}}\right)  \right. \\
&\left. - 17 x^{\frac{3}{17}} \, _2F_1\left(\frac{1}{10}, \frac{2}{3}, \frac{11}{10}, x^{\frac{30}{17}}\right)- 15 x^{\frac{5}{17}} \, _2F_1\left(\frac{1}{6}, \frac{2}{3}, \frac{7}{6}, x^{\frac{30}{17}}\right) - 14 x^{\frac{6}{17}} \, _2F_1\left(\frac{1}{5}, \frac{2}{3}, \frac{6}{5}, x^{\frac{30}{17}}\right) \right. \\
&\left. - 12 x^{\frac{8}{17}} \, _2F_1\left(\frac{4}{15}, \frac{2}{3}, \frac{19}{15}, x^{\frac{30}{17}}\right) - 11 x^{\frac{9}{17}} \, _2F_1\left(\frac{3}{10}, \frac{2}{3}, \frac{13}{10}, x^{\frac{30}{17}}\right)- 9 x^{\frac{11}{17}} \, _2F_1\left(\frac{11}{30}, \frac{2}{3}, \frac{41}{30}, x^{\frac{30}{17}}\right) \right. \\
&\left.  - 8 x^{\frac{12}{17}} \, _2F_1\left(\frac{2}{5}, \frac{2}{3}, \frac{7}{5}, x^{\frac{30}{17}}\right)- 6 x^{\frac{14}{17}} \, _2F_1\left(\frac{7}{15}, \frac{2}{3}, \frac{22}{15}, x^{\frac{30}{17}}\right) - 5 x^{\frac{15}{17}} \, _2F_1\left(\frac{1}{2}, \frac{2}{3}, \frac{3}{2}, x^{\frac{30}{17}}\right) \right. \\
&\left. - 3 x \, _2F_1\left(\frac{17}{30}, \frac{2}{3}, \frac{47}{30}, x^{\frac{30}{17}}\right) - 2 x^{\frac{18}{17}} \, _2F_1\left(\frac{3}{5}, \frac{2}{3}, \frac{8}{5}, x^{\frac{30}{17}}\right)+ x^{\frac{21}{17}} \, _2F_1\left(\frac{2}{3}, \frac{7}{10}, \frac{17}{10}, x^{\frac{30}{17}}\right)  \right. \\
&\left. - \frac{81}{23} x^{\frac{23}{17}} \, _2F_1\left(\frac{2}{3}, \frac{23}{30}, \frac{53}{30}, x^{\frac{30}{17}}\right)+ 4 x^{\frac{24}{17}} \, _2F_1\left(\frac{2}{3}, \frac{4}{5}, \frac{9}{5}, x^{\frac{30}{17}}\right) - \frac{72}{13} x^{\frac{26}{17}} \, _2F_1\left(\frac{2}{3}, \frac{13}{15}, \frac{28}{15}, x^{\frac{30}{17}}\right) \right. \\
&\left. + 7 x^{\frac{27}{17}} \, _2F_1\left(\frac{2}{3}, \frac{9}{10}, \frac{19}{10}, x^{\frac{30}{17}}\right) - \frac{189}{29} x^{\frac{29}{17}} \, _2F_1\left(\frac{2}{3}, \frac{29}{30}, \frac{59}{30}, x^{\frac{30}{17}}\right) \right].
\end{align*}
We require $D_1=1.6$.
\begin{figure}[H]
\subfigure[]{\label{fig:TeffX} %% label for second subfigure
\includegraphics[width=8.1cm]{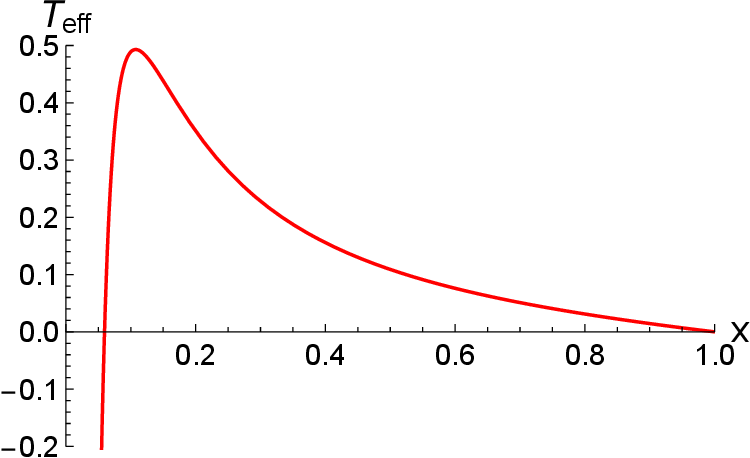}}%\\
 %\hfill
\quad
\subfigure[]{\label{fig:Sx} %% label for second subfigure
\includegraphics[width=8.1cm]{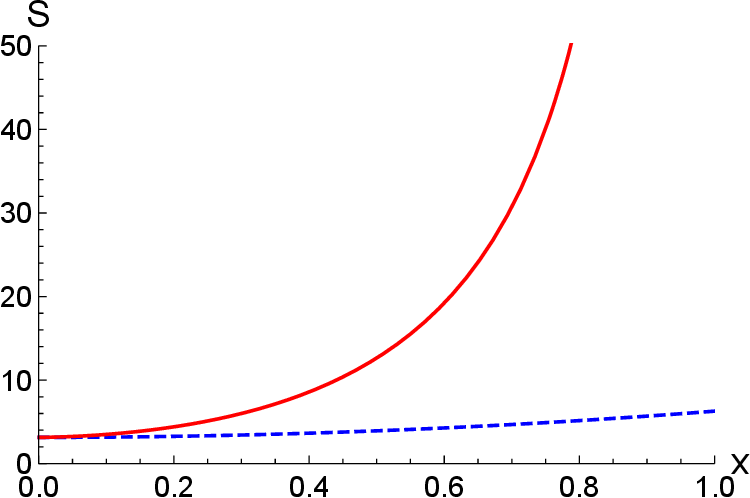}}
 %  \quad
%  \subfigure[]{\label{fig:Cpeffx} %% label for second subfigure
%  \includegraphics[width=5.3cm]{CPeff-x.eps}}
  \caption{$T_{eff}$ and total entropy $S$ with respect to $x$. In (a), $T_{eff}$ has a maximum at $x=0.107593$. In (b), the dashed (blue) curve represents the sum of the two horizon entropies and the solid (red) curve depicts the result in Eq. (\ref{eS}). 
  We set $r_c = 1$ and $Q=0.2$. }
  \label{fig:Fx}
\end{figure}

In Fig. \ref{fig:Fx}, we depict the total entropy $S$, effective temperature $T_{eff}$ as
functions of $x$. It is shown that $T_{eff}$ tends to zero as $x \rightarrow 1$, namely the charged Nariai limit. Although this result does not agree with that of Bousso and Hawking \cite{Bousso:1996}, it is consistent with the entropy. As is depicted in Fig. \ref{fig:Sx}, the entropy will diverge as  $x \rightarrow 1$. Besides, one can see that the entropy is monotonically increasing with the
increase of $x$, while $T_{eff}$ first increases and then decreases. According to the general definition of heat capacity, $C = \frac{\partial M}{\partial T} = T \frac{\partial S}{\partial T}$, these black holes can be thermodynamically stable only in the region of $x$ with the positive temperature and positive slope.
This is unexpected. It means that when the black hole horizon and the cosmological horizon are too far apart (small $x$) or too close together (large $x$), the black hole cannot be thermodynamically stable.

\section{CONCLUSIONS and DISCUSSIONS}\label{4s}

Considering the non-minimal coupling between the graviton and dilaton field, we discussed the charged massive Einstein-dilaton gravity.
According to the gravitational, dilaton, and Maxwell field equations obtained by varying the action, we derived the static spherically symmetric solutions of charged dilatonic black holes in four-dimensional spacetime.
It should be noted that for a real $\alpha$, it is required that $m_0^2 c_0 c_1 > 0$ or $m_0^2 c_0 c_1 < -1$, thereby excluding the range $-1 < m_0^2 c_0 c_1 < 0$.
Our results reduce to the Schwarzschild case when $c_1 = c_2 = Q = 0$, and reduce to the Reissner-Nordstr\"{o}m case when $c_1 = c_2 = 0$.

Later, we have analyzed the singularity of the solution,
we give the Ricci and Kretschmann scalars,
which suggest that the horizons of black holes are just singularity of coordinate as it should be.
What is more, both scalars exhibit the same asymptotic behavior at the origin.
When $m_0 = c_0 = 1$, the range $-1 < c_1 < 0$ is excluded.
For $c_1 < -1$ or $c_1 > 0$, it can be deduced that there is an essential point located at the origin, and the metric function $f(r)$ tends to infinity as $r \to \infty$.
We also show that the  black hole solutions can provide one horizon, two horizons (the inner and outer
event horizons), three horizons (the inner and outer event horizons, and the cosmological horizon), extreme (Nariai) and naked singularity black holes for the suitably fixed parameters. With dilaton field, the parameter $c_1$ affects the behavior of the metric function.

Finally, we analyzed the thermodynamics of black holes where $f(r)$  approaches $+\infty$ and $-\infty$, respectively.
In the $f(r) \rightarrow +\infty$ case, we studied the mass, temperature, and entropy of these charged dilatonic black holes, and checked the first law of black hole thermodynamics.
The analysis of the mass suggest that there could exists a minimum of the mass function of black hole horizon. For the temperature of these black holes, we found that it has a maximum positive value $T_{max}$, distinguishing two sizes of black holes (a smaller one and a larger one) with the same temperature.
Moreover, the maximal points of the temperature function correspond to the discontinuous points of the heat capacity. The domain of smaller black hole radii ($ r_{min}< r_h< r_{cri}$) demonstrates stability, while black holes with relatively large horizon radii ($r_h > r_{cri}$) are unstable.
In the $f(r) \rightarrow -\infty$ case, we have presented the entropy. It is not only the sum of the entropies of black hole horizon and the cosmological horizon, but also with an extra term from the correlation between the two horizons. 
This idea has twofold advantages. Firstly, without the additional term in the total entropy, the effective temperature is not the same as that of the black hole horizon and the cosmological horizon in the lukewarm case. Secondly, the method of effective first law of thermodynamics lacks physical explanation or motivation. While taking advantage of the method, we obtain the corrected entropy of the black hole, which may make the method more acceptable.

{\bf Acknowledgements}

Q. Y. Pan is supported by National Natural Science Foundation of China (Grant Nos. 12275079 and 12035005).
M. Zhang is supported by the Natural Science Basic Research Program of Shaanxi, China (2023-JC-QN-0053).
D. C. Zou is supported by National Natural Science Foundation of China  (Grant No. 12365009) and Natural Science Foundation of Jiangxi Province (No. 20232BAB201039).


\begin{thebibliography}{}
%\input{thebibliography}

%\input{thebibliography}



%\cite{wenberg:1989}
\bibitem{weinberg:1989}
S. Weinberg,
%``The cosmological constant problem'',
Rev.\ mod.\ phys.\ {\bf61}, 1 (1989).

%\cite{Riess:1998}
\bibitem{Riess:1998}
A. G. Riess, \textsl{et\ al.} (Supernova Search Team),
%``Observational evidence from supernovae for an accelerating universe and a cosmological constant'',
Astron.\ J\ {\bf116}, 1009 (1998).\
%\cite{Perlmutter:1999}
\bibitem{Perlmutter:1999}
Perlmutter, \textsl{ et\ al.}(Surpernova Cosmoly Project),
%Measurements of $\Omega$ and $\Lambda$ from 42 high-redshift supernovae,
Astrophys.\ J\ {\bf517}, 565 (1999).

%\cite{Planck:2015fie}
\bibitem{Planck:2015fie}
P. A. R. Ade \textit{et al.} [Planck],
%``Planck 2015 results. XIII. Cosmological parameters,''
Astron. Astrophys. \textbf{594}, A13 (2016)
%doi:10.1051/0004-6361/201525830
[arXiv:1502.01589 [astro-ph.CO]].\
%11534 citations counted in INSPIRE as of 15 Apr 2023
%\cite{WMAP:2003elm}
\bibitem{WMAP:2003elm}
D. N. Spergel \textit{et al.} [WMAP],
%``First year Wilkinson Microwave Anisotropy Probe (WMAP) observations: Determination of cosmological parameters,''
Astrophys. J. Suppl. \textbf{148}, 175-194 (2003)
%doi:10.1086/377226
[arXiv:astro-ph/0302209 [astro-ph]].
%9648 citations counted in INSPIRE as of 15 Apr 2023
%dilaton gravity
%\cite{Green:1987sw}
\bibitem{Green:1987sw}
M. B. Green, J. H. Schwarz, and E. Witten,
Superstring Theory, (Cambridge University Press, Cambridge 1987).

%\cite{Mignemi:1991wa}
\bibitem{Mignemi:1991wa}
S. Mignemi and D. L. Wiltshire,
%``Black holes in higher derivative gravity theories,''
Phys. Rev. D \textbf{46}, 1475-1506 (1992)
%doi:10.1103/PhysRevD.46.1475
[arXiv:hep-th/9202031 [hep-th]].\
%118 citations counted in INSPIRE as of 13 Apr 2023
%\cite{Poletti:1994ff}
\bibitem{Poletti:1994ff}
S. J. Poletti and D. L. Wiltshire,
%``The Global properties of static spherically symmetric charged dilaton space-times with a Liouville potential,''
Phys. Rev. D \textbf{50}, 7260-7270 (1994)
[erratum: Phys. Rev. D \textbf{52}, 3753-3754 (1995)]
%doi:10.1103/PhysRevD.50.7260
[arXiv:gr-qc/9407021 [gr-qc]].\
%127 citations counted in INSPIRE as of 13 Apr 2023
%\cite{Poletti:1994ww}

\bibitem{Poletti:1994ww}
S. J. Poletti, J. Twamley, and D. L. Wiltshire,
%``Charged dilaton black holes with a cosmological constant,''
Phys. Rev. D \textbf{51}, 5720-5724 (1995)
%doi:10.1103/PhysRevD.51.5720
[arXiv:hep-th/9412076 [hep-th]].\
%\cite{Cai:1997ii}
\bibitem{Cai:1997ii}
R. G. Cai, J. Y. Ji, and K. S. Soh,
%``Topological dilaton black holes,''
Phys. Rev. D \textbf{57}, 6547-6550 (1998)
%doi:10.1103/PhysRevD.57.6547
[arXiv:gr-qc/9708063 [gr-qc]].\
%172 citations counted in INSPIRE as of 13 Apr 2023
%\cite{Clement:2002mb}
\bibitem{Clement:2002mb}
G. Clement, D. Gal'tsov, and C. Leygnac,
%``Linear dilaton black holes,''
Phys. Rev. D \textbf{67}, 024012 (2003)
%doi:10.1103/PhysRevD.67.024012
[arXiv:hep-th/0208225 [hep-th]].
%116 citations counted in INSPIRE as of 13 Apr 2023

%\cite{Dehghani:2004sa}
\bibitem{Dehghani:2004sa}
M. H. Dehghani and N. Farhangkhah,
%``Charged rotating dilaton black strings,''
Phys. Rev. D \textbf{71}, 044008 (2005)
%doi:10.1103/PhysRevD.71.044008
[arXiv:hep-th/0412049 [hep-th]].
%45 citations counted in INSPIRE as of 29 Sep 2023

%\cite{Sheykhi:2007wg}
\bibitem{Sheykhi:2007wg}
A. Sheykhi,
%``Thermodynamics of charged topological dilaton black holes,''
Phys. Rev. D \textbf{76}, 124025 (2007)
%doi:10.1103/PhysRevD.76.124025
[arXiv:0709.3619 [hep-th]].
%80 citations counted in INSPIRE as of 13 Apr 2023
%\cite{Dehghani:2007ff}
%\cite{Gao:2004tu}
\bibitem{Gao:2004tu}
C. J. Gao and S. N. Zhang,
%``Dilaton black holes in de Sitter or Anti-de Sitter universe,''
Phys. Rev. D \textbf{70}, 124019 (2004)
%doi:10.1103/PhysRevD.70.124019
[arXiv:hep-th/0411104 [hep-th]].\
%88 citations counted in INSPIRE as of 14 Apr 2023
%\cite{Gao:2004tv}
\bibitem{Gao:2004tv}
C. J. Gao and S. N. Zhang,
%``Higher dimensional dilaton black holes with cosmological constant,''
Phys. Lett. B \textbf{605}, 185-189 (2005)
%doi:10.1016/j.physletb.2004.11.030
[arXiv:hep-th/0411105 [hep-th]].\
%58 citations counted in INSPIRE as of 14 Apr 2023


%%%%%%%%
%EDG
%\cite{Berti:2015itd}
\bibitem{Berti:2015itd}
E. Berti, E. Barausse, V. Cardoso, L. Gualtieri, P. Pani, U. Sperhake, L. C. Stein, N. Wex, K. Yagi, and T. Baker, \textit{et al.}
%``Testing General Relativity with Present and Future Astrophysical Observations,''
Class. Quant. Grav. \textbf{32}, 243001 (2015)
%doi:10.1088/0264-9381/32/24/243001
[arXiv:1501.07274 [gr-qc]].
%1105 citations counted in INSPIRE as of 26 Aug 2023

%\cite{Pani:2011xm}
\bibitem{Pani:2011xm}
P. Pani, E. Berti, V. Cardoso, and J. Read,
%``Compact stars in alternative theories of gravity. Einstein-Dilaton-Gauss-Bonnet gravity,''
Phys. Rev. D \textbf{84}, 104035 (2011)
[arXiv:1109.0928 [gr-qc]].
%137 citations counted in INSPIRE as of 26 Aug 2023
%\cite{Yunes:2011we}
\bibitem{Yunes:2011we}
N. Yunes and L. C. Stein,
%``Non-Spinning Black Holes in Alternative Theories of Gravity,''
Phys. Rev. D \textbf{83}, 104002 (2011)
[arXiv:1101.2921 [gr-qc]].
%203 citations counted in INSPIRE as of 26 Aug 2023
%\cite{Kanti:1995vq}
\bibitem{Kanti:1995vq}
P. Kanti, N. E. Mavromatos, J. Rizos, K. Tamvakis, and E. Winstanley,
%``Dilatonic black holes in higher curvature string gravity,''
Phys. Rev. D \textbf{54}, 5049-5058 (1996)
[arXiv:hep-th/9511071 [hep-th]].
%539 citations counted in INSPIRE as of 26 Aug 2023
%\cite{Torii:1996yi}
\bibitem{Torii:1996yi}
T. Torii, H. Yajima, and K. Maeda,
%``Dilatonic black holes with Gauss-Bonnet term,''
Phys. Rev. D \textbf{55}, 739-753 (1997)
[arXiv:gr-qc/9606034 [gr-qc]].
%223 citations counted in INSPIRE as of 26 Aug 2023
%\cite{Guo:2008hf}
\bibitem{Guo:2008hf}
Z. K. Guo, N. Ohta, and T. Torii,
%``Black Holes in the Dilatonic Einstein-Gauss-Bonnet Theory in Various Dimensions. I. Asymptotically Flat Black Holes,''
Prog. Theor. Phys. \textbf{120}, 581-607 (2008)
%doi:10.1143/PTP.120.581
[arXiv:0806.2481 [gr-qc]].
%93 citations counted in INSPIRE as of 26 Aug 2023
%\cite{Ohta:2009tb}
\bibitem{Ohta:2009tb}
N.~Ohta and T.~Torii,
%``Black Holes in the Dilatonic Einstein-Gauss-Bonnet Theory in Various Dimensions. III. Asymptotically AdS Black Holes with k = +-1,''
Prog. Theor. Phys. \textbf{121}, 959-981 (2009)
%doi:10.1143/PTP.121.959
[arXiv:0902.4072 [hep-th]].
%38 citations counted in INSPIRE as of 26 Aug 2023
%\cite{Ohta:2009pe}
\bibitem{Ohta:2009pe}
N.~Ohta and T.~Torii,
%``Black Holes in the Dilatonic Einstein-Gauss-Bonnet Theory in Various Dimensions IV: Topological Black Holes with and without Cosmological Term,''
Prog. Theor. Phys. \textbf{122}, 1477-1500 (2009)
%doi:10.1143/PTP.122.1477
[arXiv:0908.3918 [hep-th]].
%28 citations counted in INSPIRE as of 26 Aug 2023

%\cite{Kleihaus:2011tg}
\bibitem{Kleihaus:2011tg}
B. Kleihaus, J. Kunz, and E. Radu,
%``Rotating Black Holes in Dilatonic Einstein-Gauss-Bonnet Theory,''
Phys. Rev. Lett. \textbf{106}, 151104 (2011)
%doi:10.1103/PhysRevLett.106.151104
[arXiv:1101.2868 [gr-qc]].
%223 citations counted in INSPIRE as of 26 Aug 2023

%\cite{Maselli:2015tta}
\bibitem{Maselli:2015tta}
A. Maselli, P. Pani, L. Gualtieri, and V. Ferrari,
%``Rotating black holes in Einstein-Dilaton-Gauss-Bonnet gravity with finite coupling,''
Phys. Rev. D \textbf{92}, 083014 (2015)
%doi:10.1103/PhysRevD.92.083014
[arXiv:1507.00680 [gr-qc]].
%143 citations counted in INSPIRE as of 26 Aug 2023

%\cite{Kanti:2011jz}
\bibitem{Kanti:2011jz}
P. Kanti, B. Kleihaus, and J. Kunz,
%``Wormholes in Dilatonic Einstein-Gauss-Bonnet Theory,''
Phys. Rev. Lett. \textbf{107}, 271101 (2011)
%doi:10.1103/PhysRevLett.107.271101
[arXiv:1108.3003 [gr-qc]].
%217 citations counted in INSPIRE as of 26 Aug 2023
%\cite{Kleihaus:2016dui}
\bibitem{Kleihaus:2016dui}
B. Kleihaus, J. Kunz, S. Mojica, and M. Zagermann,
%``Rapidly Rotating Neutron Stars in Dilatonic Einstein-Gauss-Bonnet Theory,''
Phys. Rev. D \textbf{93}, 064077 (2016)
[arXiv:1601.05583 [gr-qc]].
%47 citations counted in INSPIRE as of 26 Aug 2023
%%%%%%%%%%%%%%%%%%%%%%%%%%%%%%%%%%%%%%%%%%
%\cite{Weinberg:1965pg}
\bibitem{Weinberg:1965pg}
S. Weinberg,
%``Photons and gravitons in perturbation theory: Derivation of Maxwell's and Einstein's equations'',
Phys.\ Rev.\ B\ {\bf138}, 988 (1965).\
%\cite{Boulware:1975cg}
\bibitem{Boulware:1975cg}
D. G. Boulware and S. Deser,
%``Classical general relativity derived from quantum gravity'',
Ann.\ Phys.\ {\bf89}, 193 (1975).%-240.

%\cite{Fierz:1939}
\bibitem{Fierz:1939}
M. Fierz and W. Pauli,
%``On relativistic wave equations for particles of arbitrary spin in an electromagnetic field'',
Proc.\ R.\ Soc.\ Lond.\ A\ {\bf173}, 211 (1939).%-232.

%\cite{Boulware:1972if:Boulware:1972fr}
\bibitem{Boulware:1972if}
D. G. Boulware and S. Desser,
%``Inconsistency of finite range gravitation'',
Phys.\ Lett.\ B\ {\bf40}, 227 (1972).\ %-229.
%\cite{Boulware:1972fr}
\bibitem{Boulware:1972fr}
D. G. Boulware and S. Deser,
%Can gravitation have a finite range?,
Phys.\ Rev.\ D\ {\bf6}, 3368 (1972).
%\cite{Rham:2011tl}
\bibitem{Rham:2011tl}
C. de Rham, G. Gabadadze, and A. J. Tolley,
%``Resummation of massive gravity'',
Phys.\ Rev.\ Lett.\ {\bf106}, 231101 (2011).\
%\cite{Hinterbichler:2012}
\bibitem{Hinterbichler:2012}
K. Hinterbichler,
%``Theoretical aspects of massive gravity'',
Rev.\ Mod.\ Phys.\  {\bf 84}, 671 (2012).\
%\cite{Rham:2014mg}
\bibitem{Rham:2014mg}
C. de Rham,
%``Massive gravity'',
Living\ Rev.\ Relativ.\ {\bf17}, 7 (2014).

%black hole solutions mg-bh
%\cite{Vegh:2013}
\bibitem{Vegh:2013}
D. Vegh,
%``Holography without translational symmetry'',
arXiv:\ 1301.0537\ [hep-th].\
%cite{Babichev:2014ac}
\bibitem{Babichev:2014ac}
E. Babichev and A. Fabbri,
%``A class of charged black hole solutions in massive (bi) gravity'',
J\ High\ Energy\ Phys.\, {\bf2014}, 16 (2014).
%\cite{Ghosh:2015cva}
\bibitem{Ghosh:2015cva}
S. G. Ghosh, L. Tannukij, and P. Wongjun,
%``A class of black holes in dRGT massive gravity and their thermodynamical properties,''
Eur. Phys. J. C \textbf{76}, 119 (2016)
[arXiv:1506.07119 [gr-qc]].
%thermodynamics mg-bh
%\cite{Cai:2015tb}
\bibitem{Cai:2015tb}
R. G. Cai,\ Y. P. Hu,\ Q. Y. Pan,\ and Y. L. Zhang,\
%``Thermodynamics of black holes in massive gravity'',
Phys.\ Rev.\ D\ {\bf91}, 024032 (2015)
[arXiv:1409.2369 [hep-th]].
%\cite{Xu:2015rfa}
\bibitem{Xu:2015rfa}
J. Xu, L. M. Cao, and Y. P. Hu,
%``P-V criticality in the extended phase space of black holes in massive gravity,''
Phys. Rev. D \textbf{91}, 124033 (2015)
[arXiv:1506.03578 [gr-qc]].
%209 citations counted in INSPIRE as of 30 Dec 2023
\bibitem{Alfredo:2021}
H. A. Alfredo, D. H. Daniel, and A. M. Julio, 
%``Scalarization-like mechanism through spacetime anisotropic scaling symmetry,''
Phys. Rev. D \textbf{103}, 124025 (2021)
[arXiv:2012.13412v2 [hep-th]].\

%\cite{Alfredo:2021}
\bibitem{Zangeneh:2017}
M. K. Zangeneh, B. Wang, A. Sheykhi, and Z. Y.  Tang, 
%``Charged scalar quasi-normal modes for linearly charged dilaton-Lifshitz solutions,''
Phys. Lett. B \textbf{771}, 257-263 (2017).\

%\cite{Zou:2016sab}
\bibitem{Zou:2016sab}
D. C. Zou, R. H. Yue, and M. Zhang,
%``Reentrant phase transitions of higher-dimensional AdS black holes in dRGT massive gravity,''
Eur. Phys. J. C \textbf{77}, 256 (2017) 
[arXiv:1612.08056 [gr-qc]].
%\cite{Zou:2017juz}
\bibitem{Zou:2017juz}
D. C. Zou, Y. Liu, and R. H. Yue,
%``Behavior of quasinormal modes and Van der Waals-like phase transition of charged AdS black holes in massive gravity,''
Eur. Phys. J. C \textbf{77}, 365 (2017) 
[arXiv:1702.08118 [gr-qc]].

% other bhs

\bibitem{Babichev:2014rb}
E. Babichev\ and A. Fabbri,\
%``Rotating black holes in massive gravity'',
Phys.\ Rev.\ D\ {\bf90}, 084019 (2014).\
%arXiv:1406.6096(2014).
%cite{Hendi:2015eb}
\bibitem{Hendi:2015eb}
S. H. Hendi,\ B. E. Panah,\ and S. Panahiyan,\
%``Einstein-Born-Infeld-massive gravity: adS-black hole solutions and their thermodynamical properties'',
J.\ High\ Energy\ Phys.\ {\bf2015}, 157 (2015).\
%\cite{Zhang:2017lhl}
\bibitem{Zhang:2017lhl}
M. Zhang, D. C. Zou, and R. H. Yue,
%``Reentrant phase transitions and triple points of topological AdS black holes in Born-Infeld-massive gravity,''
Adv. High Energy Phys. \textbf{2017}, 3819246 (2017)
[arXiv:1707.04101 [hep-th]].
\bibitem{Acena:2018is}
A. Ace\~{n}a,\ E. L\'{o}pez,\ and M. Llerena,\
%``Isoperimetric surfaces and area-angular momentum inequality in a rotating black hole in new massive gravity'',
Phys.\ Rev.\ D\ {\bf97}, 064043 (2018).

%cite{Kodama:2014ss}
\bibitem{Kodama:2014ss}
H. Kodama\ and I. Arraut,\
%``Stability of the Schwarzschild-de Sitter black hole in the dRGT massive gravity theory'',
Prog.\ Theor.\ Exp.\ Phys.\ {\bf2014}, 2 (2014).

\bibitem{Hendi:2016jx}
S. H. Hendi,\ G. Q. Li,\ J. X. Mo,\ S. Panahiyan,\ and B. E. Panah,\
%``New perspective for black hole thermodynamics in Gauss-Bonnet-Born-Infeld massive gravity'',
Eur.\ Phys.\ J.\  C\ {\bf76}, 571 (2016).\
\bibitem{Hendi:2018td}
S. H. Hendi\ and M. Momennia,\ J.\ High\ Energy\ Phys.\ {\bf2019}, 10 (2019).
%``Thermodynamic description of (a)dS black holes in Born-Infeld massive gravity with a non-abelian hair'',\

\bibitem{Hendi:2018gb}
S. H. Hendi,\ B. Eslam Panah,\ and S. Panahiyan,\
%``Black Hole Solutions in Gauss-Bonnet-Massive Gravity in the Presence of Power-Maxwell Field'',
Fortschr.\ Phys.\ {\bf66}, 1800005 (2018).

\bibitem{Liu:2024}
B. Liu,  R. H. Yue, D. C. Zou, L. Zhang, Z. Y. Yang,  and  Q. Pan, Phys. Rev. D {\bf109}, 064013  (2024).

\bibitem{Rakhmanov:1994}
M. Rakhmanov, Phys. Rev. D {\bf50}, 5155 (1994).

\bibitem{Poletti:1995}
S. J. Poletti, J. Twamley, and D. L. Wiltshire,
Phys. Rev. D {\bf51}, 5720  (1995).

\bibitem{Gao:2004}
C. J. Gao and S. N. Zhang, Phys. Rev. D {\bf70}, 124019 (2004).

\bibitem{Hendi:2016MF}
S. H. Hendi, M. Faizal, B. E. Panah, and S. Panahiyan, Eur. Phys. J. C {\bf76}, 296 (2016).

% dilaton massive gravity
%\cite{DAmico:2012hia}
\bibitem{DAmico:2012hia}
G. D'Amico, G. Gabadadze, L. Hui, and D. Pirtskhalava,
%``Quasidilaton: Theory and cosmology,''
Phys. Rev. D \textbf{87}, 064037 (2013)
%doi:10.1103/PhysRevD.87.064037
[arXiv:1206.4253 [hep-th]].\
%149 citations counted in INSPIRE as of 15 Apr 2023
%\cite{Gannouji:2013rwa}
\bibitem{Gannouji:2013rwa}
R. Gannouji, M. W. Hossain, M. Sami, and E. N. Saridakis,
%``Quasidilaton nonlinear massive gravity: Investigations of background cosmological dynamics,''
Phys. Rev. D \textbf{87}, 123536 (2013)
%doi:10.1103/PhysRevD.87.123536
[arXiv:1304.5095 [gr-qc]].\
%38 citations counted in INSPIRE as of 15 Apr 2023
%\cite{Wu:2016jfw}

\bibitem{Wu:2016jfw}
D. J. Wu and S. Y. Zhou,
%``No hair theorem in quasi-dilaton massive gravity,''
Phys. Lett. B \textbf{757}, 324-329 (2016)
%doi:10.1016/j.physletb.2016.04.016
[arXiv:1601.04399 [hep-th]].\
%5 citations counted in INSPIRE as of 15 Apr 2023
%\cite{Martinovic:2019hpo}
\bibitem{Martinovic:2019hpo}
K. Martinovic and M. Sakellariadou,
%``Constraints on Quasi-dilaton Massive Gravity,''
Phys. Rev. D \textbf{100}, 124016 (2019)
%doi:10.1103/PhysRevD.100.124016
[arXiv:1908.08247 [gr-qc]].\
%3 citations counted in INSPIRE as of 15 Apr 2023
%\cite{Akbarieh:2021vhv}
\bibitem{Akbarieh:2021vhv}
A. R. Akbarieh, S. Kazempour, and L. Shao,
%``Cosmological perturbations in Gauss-Bonnet quasi-dilaton massive gravity,''
Phys. Rev. D \textbf{103}, 123518 (2021)
%doi:10.1103/PhysRevD.103.123518
[arXiv:2105.03744 [gr-qc]].\
%16 citations counted in INSPIRE as of 15 Apr 2023

 %\cite{Abbott:1982}
 %\bibitem{Abbott:1982}
 %L. F. Abbott and S. Deser,\ \
 %Nucl.\ phys.\ B \textbf{195},\ 76 (1982).
\bibitem{Brown:1993}
 J. D. Brown and J. W. York, Phys. Rev. D \textbf{47}, 1407 (1993).
\bibitem{Brown:1994}
J. D. Brown, J. Creighton, and R. B. Mann, Phys. Rev. D \textbf{50}, 6394 (1994).
\bibitem{Chan:1995}
K. C. K Chan, J. H. Horne, and R. B. Mann, Nucl. Phys.
B \textbf{447}, 441 (1995).
\bibitem{Dehghani:2019}
M. Dehghani and M. R. Setare, Phys. Rev. D \textbf{100}, 044022 (2019).

\bibitem{Cai:2002}
 R. G. Cai, Nucl.\ phys.\ B \textbf{628}, 375 (2002).

\bibitem{Ma:2015}
M. S. Ma, L. C. Zhang, H. H. Zhao, and R. Zhao. Adv. High Energy Phys. \textbf{2015}, 134815 (2015).
 
\bibitem{Zhang:2019}
Y. Zhang, L. C. Zhang, and R. Zhao,  Mod. Phys. Lett. A \textbf{34}, 1950254 (2019).

\bibitem{Li:2017}
H. F. Li, M. S. Ma, L. C. Zhang, and R. Zhao, Nucl.\ phys.\ B \textbf{920}, 211 (2017).

\bibitem{Du:2022}
Y. Z. Du, H. F. Li, and L. C. Zhang, Eur. Phys. J. C \textbf{82}, 370 (2022).

\bibitem{Zhang:2016}
L. C. Zhang, R. Zhao, and M. S. Ma, Phys. Lett. B \textbf{761}, 74 (2016).

\bibitem{Zhao:2021}
H. H. Zhao, L. C.  Zhang, Y. Gao, and F. Liu, Chin. Phys. C \textbf{45}, 043111 (2021).

\bibitem{Bousso:1996}
R. Bousso, S. Hawking, Phys. Rev. D \textbf{54}, 6312 (1996).

%\cite{Dehghani:2019cuf}
%\bibitem{Dehghani:2019cuf}
%M. Dehghani and M. R. Setare,
%``Dilaton black holes with power law electrodynamics,''
%Phys. Rev. D \textbf{100}, 044022 (2019)
%doi:10.1103/PhysRevD.100.044022
%[arXiv:1906.11063 [gr-qc]].
%27 citations counted in INSPIRE as of 06 Jan 2024
%\bibitem{Zhao:2010GRG}
%R. Zhao, L. C. Zhang, and H. F. Li, Gen. Relat. Gravit. \textbf{42}, 4  (2010).
%\bibitem{Zhao:2010EPJC}
%R. Zhao, L. Zhang, H. Li, and Y. Wu,  Eur. Phys. J. C \textbf{65}, 1  (2010).
%\bibitem{Zhao:2013CPMA}
%R. Zhao and L. C. Zhang,  Sci. China Phys. Mech. Astron. \textbf{56}, 9  (2013).
%\bibitem{Dolan:2014PRD}
%B. P. Dolan, Phys. Rev. D \textbf{90}, 8  (2014).
%\bibitem{Mirza:2014PRD}
%B. Mirza and Z. Sherkatghanad, Phys. Rev. D \textbf{90}, 8  (2014).
%\bibitem{Cai:2002NPB}
%R. G. Cai, Nucl. phys. B \textbf{628}, 1  (2002).
%\bibitem{Sekiwa:2006PRD}
%Y. Sekiwa, Phys. Rev. D \textbf{73}, 8  (2006).


\end{thebibliography}
\end{document}